\documentclass[%
  reprint,
 showpacs,
 amsmath,
 amssymb,
 aps,
 prc,
]{revtex4-1}

\usepackage[caption=false]{subfig}
\usepackage{tabularx}
\usepackage{graphicx}
\graphicspath{{./figs/}}
\usepackage{amsmath}
\usepackage{todonotes}
\usepackage{hyperref}

\renewcommand{\d}{\mathrm{d}}
\newcommand{\p}{\mathrm{p}}
\newcommand{\n}{\mathrm{n}}

\renewcommand{\phi}{\varphi}
\renewcommand{\rm}[1]{\textnormal{#1}}

\renewcommand{\vec}[1]{\boldsymbol{#1}}

\newcommand{\bra}[1]{\langle \, #1 \,  | \,}
\newcommand{\ket}[1]{\, | \, #1 \, \rangle}

\usepackage[version=4]{mhchem}      

\def\inclebsch(#1,#2,#3,#4,#5,#6){\langle #1 \hspace{5pt} #2 \hspace{5pt} #3 \hspace{5pt} #4 \hspace{2pt} |\hspace{2pt} #5 \hspace{5pt} #6 \rangle}

\def\threej#1{\inthreej(#1)}
\def\inthreej(#1,#2,#3,#4,#5,#6){\begin{pmatrix}#1 & #2 & #3 \\ #4 & #5 & #6 \end{pmatrix}}

\def\insixj(#1,#2,#3,#4,#5,#6){\begin{Bmatrix}#1 & #2 & #3 \\ #4 & #5 & #6 \end{Bmatrix}}

\def\ninej#1{\inninej(#1)}
\def\inninej(#1,#2,#3,#4,#5,#6,#7,#8,#9){\begin{Bmatrix}#1 & #2 & #3 \\#4 & #5 & #6 \\#7 & #8 & #9 \end{Bmatrix}}

\makeatletter
\newcommand{\Vast}{\bBigg@{3}}
\makeatother

\allowdisplaybreaks

\begin{document}


\title{Seagull and pion-in-flight currents in neutrino-induced $1N$ and $2N$ knockout}


\author{T.~Van Cuyck}
\email{Tom.VanCuyck@UGent.be}
\affiliation{Department of Physics and Astronomy,\\
Ghent University,\\ Proeftuinstraat 86,\\ B-9000 Gent, Belgium}
\author{N.~Jachowicz}
\email{Natalie.Jachowicz@UGent.be}
\affiliation{Department of Physics and Astronomy,\\
Ghent University,\\ Proeftuinstraat 86,\\ B-9000 Gent, Belgium}
\author{R.~Gonz\'{a}lez-Jim\'{e}nez}
\affiliation{Department of Physics and Astronomy,\\
Ghent University,\\ Proeftuinstraat 86,\\ B-9000 Gent, Belgium}
\author{J.~Ryckebusch}
\affiliation{Department of Physics and Astronomy,\\
Ghent University,\\ Proeftuinstraat 86,\\ B-9000 Gent, Belgium}
\author{N.~Van Dessel}
\affiliation{Department of Physics and Astronomy,\\
Ghent University,\\ Proeftuinstraat 86,\\ B-9000 Gent, Belgium}



\date{\today}

\begin{abstract}
  \begin{description}  
    \item[Background] The neutrino-nucleus ($\nu A$) cross section is a major source of systematic uncertainty in neutrino-oscillation studies. A precise $\nu A$ scattering model, in which multinucleon effects are incorporated, is pivotal for an accurate interpretation of the data. 
    \item[Purpose] In $\nu A$ interactions, meson-exchange currents (MECs) can induce two-nucleon ($2N$) knockout from the target nucleus, resulting in a two-particle two-hole (2p2h) final state. They also affect single nucleon ($1N$) knockout reactions, yielding a one-particle one-hole (1p1h) final state. Both channels affect the inclusive strength. 
      We present a study of axial and vector, seagull and pion-in-flight currents in muon-neutrino induced $1N$ and $2N$ knockout reactions on \ce{^{12}C}. 
    \item[Method] Bound and emitted nucleons are described as Hartree-Fock wave functions. For the vector MECs, the standard expressions are used. For the axial current, three parameterizations are considered. The framework developed here allows for a treatment of MECs and short-range correlations (SRCs).
    \item[Results] Results are compared with electron-scattering data and with literature. The strengths of the seagull, pion-in-flight and axial currents are studied separately and double differential cross sections including MECs are compared with results including SRCs. A comparison with MiniBooNE and T2K data is presented. 
    \item[Conclusions] In the 1p1h channel, the effects of the MECs tend to cancel each other, resulting in a small effect on the double differential cross section. $2N$ knockout processes provide a small contribution to the inclusive double differential cross section, ranging from the $2N$ knockout threshold into the dip region. A fair agreement with the MiniBooNE and T2K data is reached. 
  \end{description}
\end{abstract}

\pacs{25.30.Pt,13.15.+g,24.10.Cn,25.40.-h}

\maketitle

\section{Introduction}

With the advent of accelerator-based neutrino-oscillation experiments, the precision of the determined neutrino-oscillation parameters improved a lot. A major source of systematic uncertainty in the analyses is related to the neutrino-nucleus scattering cross sections. 
To further improve the precision of the determined squared-mass differences and mixing angles, an accurate neutrino-nucleus ($\nu A$) interaction model is required. 
Progress and issues in this context have recently been reviewed in Refs.~\cite{Mosel:2016cwa,Katori:2016aaa}.
One of the main challenges is related to the role of multinucleon effects. 

In previous work, we studied the effect of long-range correlations in a continuum random-phase approximation (CRPA) approach \cite{Jachowicz:1998fn,Jachowicz:2002rr,Pandey:2013cca,Pandey:2014tza,Pandey:2016jju,Martini:2016eec} and short-range correlations (SRCs) \cite{VanCuyck:2016fab}. 
This work is a further development and focuses on the influence of the seagull and pion-in-flight currents, and accounts for one-nucleon ($1N$) and two-nucleon ($2N$) knockout interactions.

In our model, the initial and final state of the nucleus is described as a Slater determinant. 
Mean-field single particle wave functions from a Hartree-Fock (HF) calculation are used. These HF wave functions account for the elastic distortion by the residual nuclear system on the emitted nucleons.
Shell structure, nuclear binding energy and Pauli-blocking are included. 
The model is an extension towards the weak sector of the $2N$ knockout model developed in Ghent, which accounts for meson-exchange currents (MECs), $\Delta$-currents as well as SRCs, for photoinduced \cite{Ryckebusch:1993tf} and electroinduced \cite{Ryckebusch:1997gn,Janssen:1999xy} $1N$ and $2N$ knockout reactions. 
The model describes exclusive $(e,e^\prime N N)$ \cite{Starink:2000aa,Ryckebusch:2003tu}, 
semi-exclusive $(e,e^\prime \textnormal{p})$ \cite{Fissum:2004we,Iodice:2007mn} 
and inclusive $(e,e^\prime)$ \cite{VanderSluys:1993cv} scattering with a satisfactory accuracy. 
The $\Delta$-currents are not included here.

Several theoretical approaches have analyzed the role of MECs in $\nu A$ interactions. 
The models by Martini \textit{et~al.}~\cite{Martini:2009uj} and Nieves \textit{et~al.}~\cite{Nieves:2011pp} take nuclear finite-size effects into account via a local density approximation and a semi-classical expansion of the response function. Both approaches include the interference between MECs, $\Delta$-currents and the correlation current. 
Recently, calculations using a relativistic Fermi gas by Amaro \textit{et al.}~\cite{Amaro:2010sd}, accounting for correlations, MECs and $\Delta$-currents in electroinduced $2N$ emission, 
have been extended to $\nu A$ and $\overline{\nu}A$ interactions \cite{Simo:2014wka,Simo:2014esa,Simo:2016ikv,Simo:2016ikw}. 
In ab-initio calculations on $^{12}$C \cite{Lovato:2013cua,Lovato:2015qka}, MECs are inherently included. 
Recent work on electron scattering \cite{Benhar:2015ula,Rocco:2015cil} has generalized the formalism based on a factorization ansatz and nuclear spectral functions to treat transition matrix elements involving MECs and $\Delta$-currents.

The structure of this work is as follows. 
In Sec.~\ref{sec:mec}, the seagull, pion-in-flight and axial currents used in the numerical calculations are discussed. 
The influence of the MECs on $1N$ emission processes is studied in Sec.~\ref{sec:one}. 
$2N$ knockout of MEC pairs is outlined in Sec.~\ref{sec:two}, where exclusive, semi-exclusive and inclusive cross sections are studied. 
In Sec.~\ref{sec:flu}, the computed $2N$ knockout strength of MECs and SRCs is added to the $1N$ knockout strength in the CRPA approach and theoretical predictions for the MiniBooNE and T2K data are provided. 
In Sec.~\ref{sec:sum}, our conclusions are presented. 

\section{Seagull and pion-in-flight currents}\label{sec:mec}

\begin{figure*}[ht]
  \centering
  \subfloat[a][]{\includegraphics[width=0.20\textwidth]{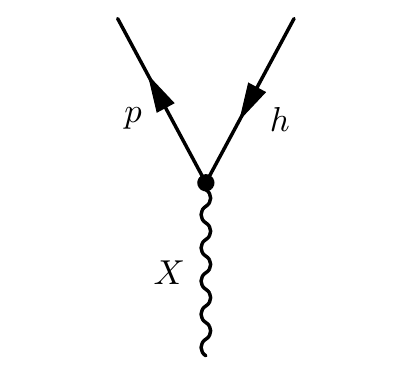}} 
  \subfloat[b][]{\includegraphics[width=0.20\textwidth]{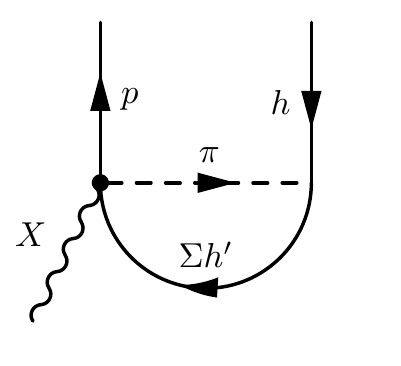}} 
  \subfloat[c][]{\includegraphics[width=0.20\textwidth]{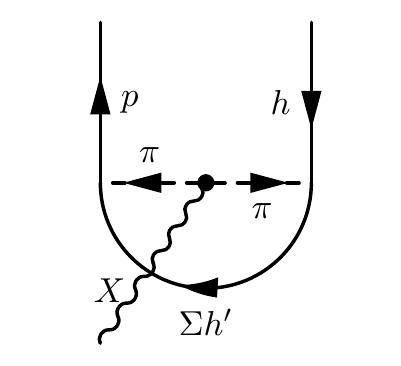}} 
  \subfloat[d][]{\includegraphics[width=0.20\textwidth]{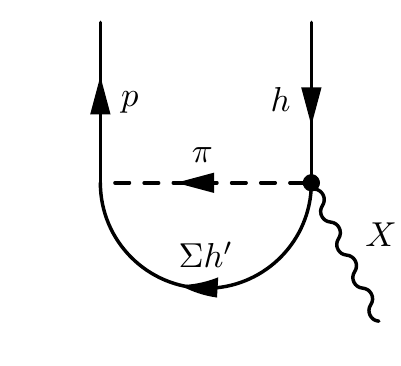}} \\
  \subfloat[e][]{\includegraphics[width=0.20\textwidth]{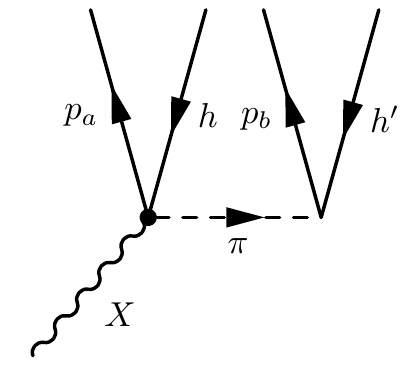}} 
  \subfloat[f][]{\includegraphics[width=0.22\textwidth]{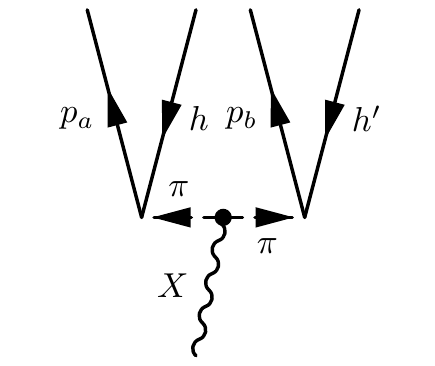}}
  \subfloat[g][]{\includegraphics[width=0.20\textwidth]{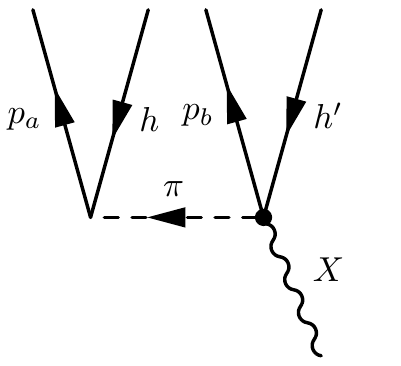}}
  \caption{The vector current diagrams considered in this paper. Diagram (a) shows the 1p1h channel in the IA, diagrams (b,d) and (e,g) are the 1p1h and 2p2h seagull diagrams and (c) and (f) the pion-in-flight diagrams.}
  \label{fig:onetwomec}
\end{figure*}

The MECs considered in this work are the seagull and pion-in-flight currents. 
The conventional approach is to consider all diagrams with single-pion exchange. 
In the seagull currents, the boson couples with the MEC at the $\pi NN$ vertex, while in pion-in-flight currents, the boson couples with the virtual pion. The vector MECs are shown in Fig.~\ref{fig:onetwomec} for the 1p1h and the 2p2h channel. The sum $\sum_{h^\prime}$ in the 1p1h channel extends over all occupied single-particle states of the target nucleus, as explained in \cite{VanCuyck:2016fab}.
In the derivation of the Feynman diagrams, the couplings are either obtained from a pion-nucleon scattering amplitude \cite{Chemtob:1971pu} or from an effective chiral Lagrangian \cite{Riska:1972zz}. 
In the low-energy limit, the vector seagull (labeled 'sea') and pion-in-flight ('pif') currents, for electron scattering interactions, are given by \cite{Towner:1987zz,Riska:1989bh,Mathiot:1989vw,Towner:1992np}
\begin{widetext}
\begin{align}
  \vec{\widehat{J}}^{\,[2],\rm{sea}}_V &= -i \left( \frac{f_{\pi NN}}{m_\pi} \right)^2 (\vec{I}_V)_3 \, F_1^V(Q^2) \left( \Gamma_\pi^2(\vec{q}_2^2) \frac{ \vec{\sigma}_1\left( \vec{\sigma}_2 \cdot \vec{q}_2 \right)}{\vec{q}_2^2+m_\pi^2} - \Gamma_\pi^2(\vec{q}_1^2)\frac{  \vec{\sigma}_2\left( \vec{\sigma}_1 \cdot \vec{q}_1 \right)}{\vec{q}_1^2+m_\pi^2} \right) \label{eq:seaform1}, \\
  \vec{\widehat{J}}^{\,[2],\rm{pif}}_V &= i\left( \frac{f_{\pi NN}}{m_\pi} \right)^2 (\vec{I}_V)_3 \, F_1^V(Q^2) F(\vec{q}_1^2,\vec{q}_2^2) 
  \frac{\left(\vec{\sigma}_1 \cdot \vec{q}_1 \right) \left(\vec{\sigma}_2 \cdot \vec{q}_2 \right)}{\left(\vec{q}_1^2+m_\pi^2\right)\left(\vec{q}_2^2+m_\pi^2 \right)} 
  (\vec{q}_1-\vec{q}_2) \label{eq:pifform1}, 
\end{align}
\end{widetext}
where $\vec{I}_V$ is the two-body isovector operator
\begin{align}
  \vec{I}_V = \left( \vec{\tau}_1 \times \vec{\tau}_2 \right).
\end{align}
The currents for a CC neutrino interaction can be obtained via an isospin rotation, which follows from conservation of the vector current (CVC). This implies replacing the third component of the isovector operator with the $\pm$ components \cite{Towner:1987zz}
\begin{align}
  (\vec{I}_V)_3  \rightarrow & (\vec{I}_V)_\pm = \frac{1}{2} \left( (\vec{I}_V)_x \pm i (\vec{I}_V)_y \right).
\end{align}
%

The value of the $\pi NN$ coupling constant is determined via $f^2_{\pi NN}/4\pi = 0.075$, and $m_\pi$ is the mass of the pion. 
The $\pi NN$ vertices are regularized by introducing a monopole form factor with cutoff mass $\Lambda_\pi = 1250$ MeV. We follow the procedure introduced in \cite{Mathiot:1989vw} to ensure CVC
\begin{align}
  \Gamma_\pi(\vec{q}^2) &= \frac{\Lambda_\pi^2 - m_\pi^2}{\vec{q}^2 + \Lambda_\pi^2}, \\
  F(\vec{q}_1^2,\vec{q}_2^2) &= \Gamma_\pi(\vec{q}_1^2) \Gamma_\pi(\vec{q}_2^2) \left( 1 + \frac{\vec{q}_1^2 + m_\pi^2}{\vec{q}_2^2 + \Lambda_\pi^2} + \frac{\vec{q}_2^2 + m_\pi^2}{ \vec{q}_1^2 + \Lambda_\pi^2} \right).
\end{align}
At the electroweak vertices, we introduce the isovector nucleon form factor $F_1^V(Q^2)$, using the conventions of \cite{VanCuyck:2016fab}. 

For MECs where a single pion is exchanged, only the seagull current has an axial counterpart. In the low energy limit it is given by \cite{Kubodera:1978wr,Towner:1992np}
\begin{align}
  \widehat{\rho}^{\,[2],\rm{sea}}_A &= \frac{i}{g_A} \left( \frac{f_{\pi NN}}{m_\pi} \right)^2 (\vec{I}_V)_\pm \, \nonumber \\ &\times \left( \frac{  \vec{\sigma}_2 \cdot \vec{q}_2 }{\vec{q}_2^2+m_\pi^2} - \frac{  \vec{\sigma}_1 \cdot \vec{q}_1 }{\vec{q}_1^2+m_\pi^2} \right),\label{eq:axseanoform} 
\end{align}
with $g_A=1.26$.
The $\pi NN$ vertex is regularized by introducing monopole factors as was done for the vector seagull current. At the electroweak vertices one relies on the partially conserved axial current (PCAC) hypothesis to constrain the currents. In the low-energy limit, this procedure is not unambiguous and different results are found in literature. 
An in-depth discussion of these differences is beyond the scope of this paper but can be found e.g.~in \cite{Towner:1987zz,Towner:1992np}. In this work, we consider three different prescriptions for the axial current. 
The first two are different parameterizations for the axial seagull current and the third expression contains more diagrams next to the axial seagull current
\begin{widetext}
\begin{align}
  \widehat{\rho}^{\,[2],\rm{sea},1}_{A} &= \frac{i}{g_A} \left( \frac{f_{\pi NN}}{m_\pi} \right)^2 (\vec{I}_V)_\pm  \, G_A(Q^2) \, \left(\Gamma_\pi^2(\vec{q}_2^2) \frac{ \vec{\sigma}_2 \cdot \vec{q}_2 }{\vec{q}_2^2+m_\pi^2} -  \Gamma_\pi^2(\vec{q}_1^2)\frac{  \vec{\sigma}_1 \cdot \vec{q}_1 }{\vec{q}_1^2+m_\pi^2} \right) \label{eq:axseaform1}, \\
  \widehat{\rho}^{\,[2],\rm{sea},2}_{A} &= \frac{i}{g_A} \left( \frac{f_{\pi NN}}{m_\pi} \right)^2 (\vec{I}_V)_\pm \, \left(F_\pi(\vec{q}_1^2) \,\Gamma_\pi^2(\vec{q}_2^2) \frac{ \vec{\sigma}_2 \cdot \vec{q}_2 }{\vec{q}_2^2+m_\pi^2} - F_\pi(\vec{q}_2^2) \,\Gamma_\pi^2(\vec{q}_1^2)\frac{  \vec{\sigma}_1 \cdot \vec{q}_1 }{\vec{q}_1^2+m_\pi^2} \right) \label{eq:axseaform2}, \\
  \widehat{\rho}^{\,[2],\rm{axi}}_{A} &= \frac{i}{g_A} \left( \frac{f_{\pi NN}}{m_\pi} \right)^2 (\vec{I}_V)_\pm \, \left(F_\pi(\vec{q}_2^2) \,\Gamma_\pi^2(\vec{q}_2^2) \frac{ \vec{\sigma}_2 \cdot \vec{q}_2 }{\vec{q}_2^2+m_\pi^2} - F_\pi(\vec{q}_1^2) \,\Gamma_\pi^2(\vec{q}_1^2)\frac{  \vec{\sigma}_1 \cdot \vec{q}_1 }{\vec{q}_1^2+m_\pi^2} \right) \label{eq:axseaform3}.
\end{align}
\end{widetext}
Pion form factors are introduced to comply with the PCAC hypothesis  
\begin{align}
  F_\pi(\vec{q}^2) &= \frac{m_\rho^2}{\vec{q}^2 + m_\rho^2}.
  \label{}
\end{align}

The current, labeled with the superscript 'sea,1', is the axial version of the seagull current, derived using the soft-pion approximation \cite{Kubodera:1978wr,Towner:1992np}. It can be constructed from Eq.~(\ref{eq:axseanoform}) by introducing the monopole form factors $\Gamma_\pi(\vec{q}_i^2)$ at the $\pi NN$ vertices and multiplying it by the axial form factor $G_A(Q^2)$, for which we adopt the standard dipole parameterization. 
The expression (\ref{eq:axseaform1}) was used in the neutrino-deuteron scattering studies of Refs.~\cite{Nakamura:2002jg} and \cite{Shen:2012xz}. 

For the construction of the axial seagull current with superscript 'sea,2', a nonrelativistic reduction of the axial seagull current used in the calculations by Ruiz Simo \textit{et al.}~\cite{Simo:2016ikv} was performed. In that work, the form factors were based on those used in the weak pion production amplitudes of \cite{Hernandez:2007qq}. The MECs were constructed by appending the pion production diagrams with an extra nucleon that absorbs the virtual pion. 
The pion form factor was introduced to account for the $\rho$-meson dominance of the $\pi \pi NN$ vertex. To account for the \textit{one-body version} of PCAC, the same form factor was used to regularize the axial $W \pi NN$ vertex.
The $\pi NN$ vertices are multiplied by the $\Gamma_\pi(\vec{q}_i^2)$ hadronic form factors as was also done for the vector currents. 
We remark that the vector currents in this work correspond with the nonrelativistic limits of the vector seagull and pion-in-flight currents of \cite{Simo:2016ikv}.

\begin{figure}[h!]
  \centering
  \subfloat[b][]{\includegraphics[width=0.20\textwidth]{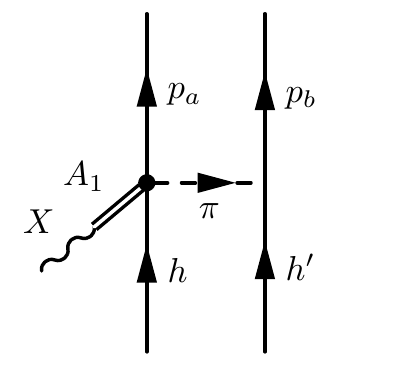}} 
  \subfloat[c][]{\includegraphics[width=0.20\textwidth]{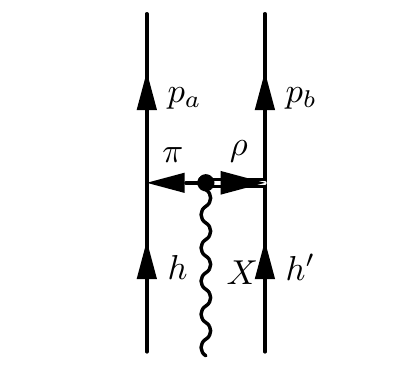}} \\
  \subfloat[d][]{\includegraphics[width=0.20\textwidth]{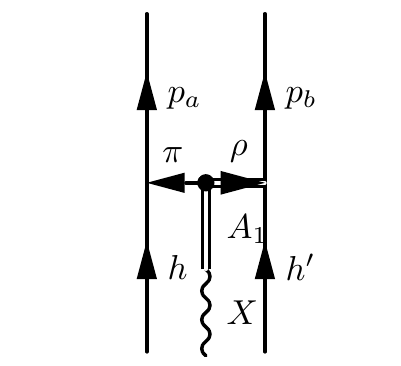}}
  \subfloat[e][]{\includegraphics[width=0.20\textwidth]{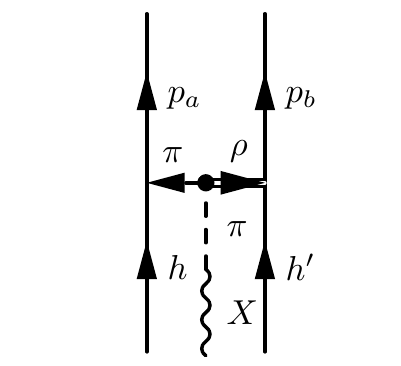}}
  \caption{Diagrams considered in the axial charge density $\widehat{\rho}_A^{\,[2],\rm{axi}}$ derived in \cite{Towner:1992np}.}\label{fig:axialdia}
 \end{figure}

 The axial current, labeled 'axi' was derived in \cite{Towner:1992np}. The four diagrams displayed in Fig.~\ref{fig:axialdia} are included.
 The first is the axial version of the seagull current (a). 
 The other three diagrams have a pion-in-flight-like structure, but one of the two pions is replaced by a $\rho$-meson, and the coupling of the $W$-boson at the $\pi \rho$ vertex is a contact coupling (b), an $A_1$-pole (c) or $\pi$-pole coupling (d). 
 The three diagrams with a $\pi-\rho$ exchange (b-d) have no vector counterpart and since one of the two mesons is a pion, they are of the same range as the vector diagrams. 
 The pion-in-flight diagrams shown in Fig.~\ref{fig:onetwomec} have no axial counterpart.
The combination of these four currents obeys the \textit{two-nucleon version} of the PCAC relation. The nonrelativistic limit of these currents is purely time-like. 
The vertices are multiplied by the appropriate $\Gamma_\pi(\vec{q}_i^2)$ form factors. 
This current has the same operator structure as the two axial seagull currents, though, by construction, it contains more diagrams. This axial current fits most naturally in our model, as it uses the two-nucleon version of the PCAC relation to constrain the currents. 

\section{MEC corrections to inclusive one-nucleon knockout}\label{sec:one}

\begin{figure*}[ht]
  \centering
    \includegraphics[width=0.95\textwidth,trim=0cm 0.6cm 0cm 0.3cm,clip]{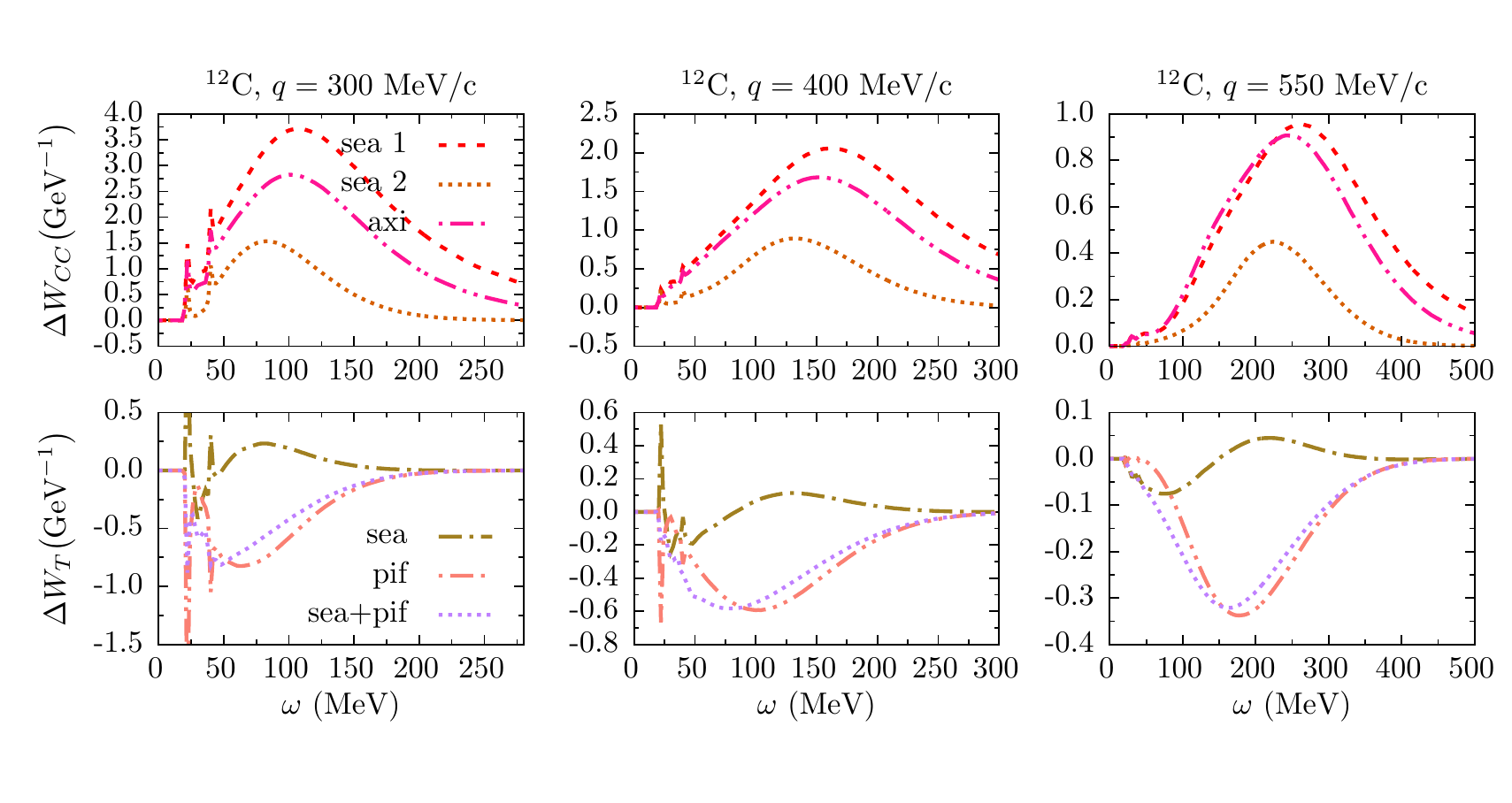}
    \caption{The correction of the MECs on the 1p1h responses for \ce{^{12}C}$(\nu_\mu,\mu^-)$ at three different $q$.} 
  \label{fig:neut1p1hqtwomec}
 \end{figure*}
  
In this section, we consider the following electron and charged-current (CC) muon-neutrino ($\nu_\mu$) induced $1N$ knockout reactions,
  \begin{align}
    e(E_e,\vec{k}_e) + A &\rightarrow e'(E_{e'},\vec{k}_{e'}) + (A-1)^* + N(E_N,\vec{p}_N) \nonumber \\
    \nu_\mu(E_{\nu_\mu},\vec{k}_{\nu_\mu}) + A &\rightarrow \mu(E_\mu,\vec{k}_\mu) + (A-1)^* + N(E_N,\vec{p}_N). \nonumber
  \end{align}
  The residual $(A-1)^*$ nucleus is left with little to no excitation energy. 
  The initial lepton will be referred to as $l$ and the final state lepton as $l'$. The four-momentum transfer, $q^\mu=(\omega,\vec{q})$, is
  \begin{align}
    \omega &= E_l - E_{l'}, & \vec{q} &= \vec{k}_l - \vec{k}_{l'},
  \end{align}
  and $Q^2 = \vec{q}^{\,2} - \omega^2$. The double differential $A(e,e^\prime)$ cross section is given by
  \begin{align}
    \dfrac{\mathrm{d}\sigma}{\mathrm{d}E_{e'} \mathrm{d}\Omega_{e'}} = \sigma^\mathrm{Mott} \bigl[& v^e_{L} W_{CC} + v^e_T W_T \bigr].\label{eq:elecone}
  \end{align}
  For $A(\nu_\mu,\mu^-)$ interactions, one has
  \begin{align}
    \dfrac{\mathrm{d}\sigma}{\mathrm{d}E_\mu \mathrm{d}\Omega_\mu} = \sigma^{W} \zeta \bigl[& v_{CC} W_{CC} + v_{CL} W_{CL} + v_{LL} W_{LL} \nonumber \\
    &+ v_T W_T  \mp v_{T'} W_{T'} \bigr],\label{eq:neutone}
  \end{align}
  the $-(+)$ sign refers to neutrino(antineutrino) scattering. The prefactors are defined as
  \begin{align}
    \sigma^\mathrm{Mott} &= \left( \frac{\alpha \cos(\theta_{e'}/2)}{2E_e\sin^2(\theta_{e'}/2)}\right)^2, \\ 
    \sigma^W &= \left( \frac{G_F \cos(\theta_c) E_\mu}{2\pi}\right)^2,
  \end{align}
  with $\alpha$ the fine-structure constant, $\theta_{e'}$ the electron scattering angle, $G_F$ the Fermi constant, $\theta_c$ the Cabibbo angle and the kinematic factor $\zeta$
  \begin{align}
    \zeta=\sqrt{1-\frac{m_\mu^2}{E_\mu^2}}.
  \end{align}
  The functions $v_i$ contain the lepton kinematics and the response functions $W_i$ the nuclear dynamics. The $W_i$ are defined as products of transition matrix elements 
  $\mathcal{J}_\lambda$
  \begin{align}
    \mathcal{J}_\lambda = \langle \Psi^{1p1h} | \widehat{J}_\lambda(q) | \Psi_{gs} \rangle. \label{eq:currents}
  \end{align}
  Here, $| \Psi^{1p1h} \rangle$ and $|\Psi_{gs} \rangle$ refer to the one-particle one-hole (1p1h) final state and the $0^+$ ground state of the target nucleus. 
  $\widehat{J}_\lambda$ are the timelike and spherical components of the nuclear current. To account for MECs, the nuclear current is written as a sum of the IA and MEC contributions
  \begin{align}
    \widehat{J}_\lambda(q) = \widehat{J}_\lambda^\textnormal{\,[1],IA}(q) + \widehat{J}_\lambda^\textnormal{\,[2],MEC}(q).
  \end{align}
  The results presented in this work consider $^{12}$C as target nucleus. For $^{12}$C$(e,e^\prime)$  two 1p1h final states are accessible 
  \begin{align}
    | \Psi^{1p1h} \rangle = | ^{11}\textnormal{C},\textnormal{n} \rangle ,\, | ^{11}\textnormal{B},\textnormal{p} \rangle,
  \end{align}
  while for CC neutrino scattering only one 1p1h final state is accessible
  \begin{align}
    | \Psi^{1p1h} \rangle = | ^{11}\textnormal{C},\textnormal{p} \rangle .
  \end{align}
  
  The expressions for the kinematic factors $v_i$ and the response functions $W_i$ can be found in \cite{VanCuyck:2016fab}. 
  The standard expressions for the nuclear current in the IA are adopted \cite{walecka2004theoretical}.
  Nucleon knockout occurs in the spectator approach (SA), where the nucleon absorbing the boson is the one that becomes asymptotically free. 
  The bound-state and continuum wave functions are constructed through a HF calculation with an effective Skyrme-type interaction \cite{Waroquier:1986mj}. Relativistic corrections are implemented in an effective fashion as explained in Refs.~\cite{Amaro:1995kk,Amaro:2005dn}.
  The wave functions for the target and residual nucleus are represented as Slater determinants. A multipole expansion is adopted for the calculation of the transition matrix elements. 

  In Fig.~\ref{fig:neut1p1hqtwomec} the difference between the 1p1h responses for  \ce{^{12}C}$(\nu_\mu,\mu^-)$, calculated with and without MECs, is shown 
  \begin{align}
    \Delta W_i = W_i^\textnormal{IA+MEC} - W_i^\textnormal{IA}.
  \end{align}
  The total 1p1h responses will be compared with the 2p2h contributions in Fig.~\ref{fig:neut1p1h2p2hqmec}.
  The three expressions for the axial current interfere constructively with the nuclear current in the IA, resulting in an increase of the Coulomb response. The effect for the 'sea,2' version is the smallest. The current 'axi' yields an increase of $\approx 10\%$ of the 1p1h Coulomb response in the IA (see Fig.~\ref{fig:neut1p1h2p2hqmec} below).
  The combined effect of the seagull and pion-in-flight currents results in a negligible decrease of the 1p1h response, the total decrease is less than $1\%$ compared to the 1p1h response in the IA. 
  In fact it is smaller than the variation obtained using alternative parameterizations of the nucleon form factor. 
  The small impact is partly due to the fact that a large part of the transverse strength comes from the axial part of the current, which is unaffected by the MECs in the low-energy limit.
    We note that the effect of the MECs on the $1N$ knockout channel of the double differential cross sections will be negligible since the cross section is dominated by the transverse channel. 

  The influence of the MECs on the 1p1h transverse response function for $^{12}$C$(e,e^\prime)$ interactions is of similar size, but has an opposite effect, increasing the response function. The reason for this opposite behavior is related to the isospin operators. 

  \section{Knockout of MEC pairs}\label{sec:two}
  For knockout of MEC pairs, we consider the following reactions
  \begin{align}
    e(E_e,\vec{k}_e) + A &\rightarrow e'(E_{e'},\vec{k}_{e'}) + (A-2)^* \nonumber \\
    &\quad + N_a(E_a,\vec{p}_a) + N_b(E_b,\vec{p}_b),\\
    \nu_\mu(E_{\nu_\mu},\vec{k}_{\nu_\mu}) + A &\rightarrow \mu(E_{\mu},\vec{k}_{\mu}) + (A-2)^* \nonumber \\
    &\quad + N_a(E_a,\vec{p}_a) + N_b(E_b,\vec{p}_b).
  \end{align}
  The residual $(A-2)^*$ nuclear system is left with little to no excitation energy. 
  Electron interactions with MECs can only emit pn pairs, due to the $(\vec{I}_V)_3$ operator, hence the 2p2h final state is
  \begin{align}
    | \Psi^{2p2h} \rangle = | ^{10}\textnormal{B},\textnormal{pn} \rangle.
  \end{align}
  For CC neutrino reactions, the pp and pn emission channels are open,
  \begin{align}
    | \Psi^{2p2h} \rangle = | ^{10}\textnormal{B},\textnormal{pp} \rangle ,\, | ^{10}\textnormal{C},\textnormal{pn} \rangle.
  \end{align}
  The two-body transition matrix elements are given by
  \begin{align}
    \mathcal{J}_\lambda = \langle \Psi^{2p2h} | \widehat{J}^\textnormal{\,[2],MEC}_\lambda(q) | \Psi_{gs} \rangle. \label{eq:twotwo}
  \end{align}
  Only the two-body part of the nuclear current contributes to the $2N$ knockout cross section. We follow the same approach as for the $1N$ knockout calculations. The SA is adopted: the pair interacting with the incoming boson is the one that becomes asymptotically free. The continuum and bound-state wave functions are calculated in the same mean-field potential. Mutual interactions between the emitted nucleons are neglected. 
The wave functions for both outgoing nucleons are expanded in terms of the continuum eigenstates of the potential and a multipole expansion is adopted for the calculation of the matrix elements \cite{Ryckebusch:1993tf}.
The 2p2h matrix elements are summarized in the appendix.
  
  \subsection{Exclusive $2N$ knockout}

  \begin{figure}[ht!]
    \centering
    \includegraphics[height=0.8\textheight]{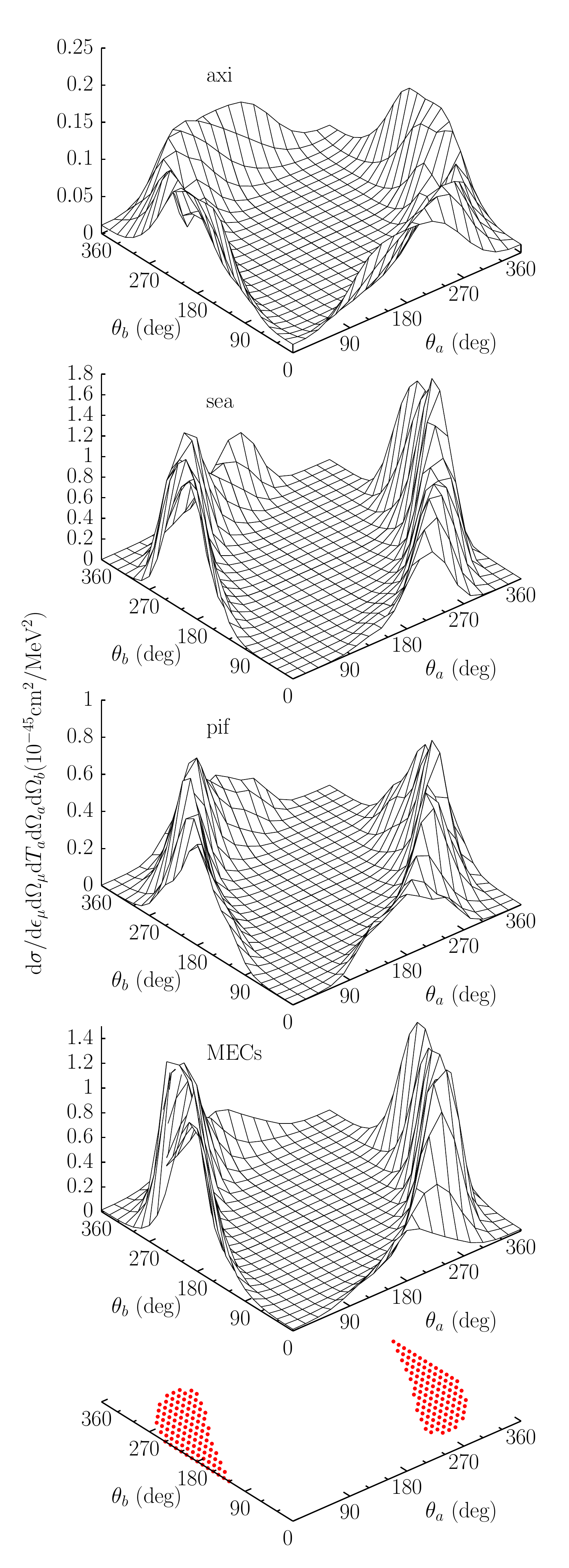}
    \caption{The $^{12}$C$(\nu_\mu,\mu^- N_a N_b)$ cross section ($N_a =$ p, $N_b = $ p$^\prime$, n) at $E_{\nu_\mu} = 750$ MeV, $E_\mu=550$ MeV, $\theta_\mu=15^\circ$ and $T_\textrm{p}=50$ MeV for in-plane kinematics. The bottom plot shows the ($\theta_a,\theta_b)$ regions with $P_{12} < 300$ MeV/c.}
    \label{fig:mecexcl}
  \end{figure}

  Exclusive $2N$ knockout refers to reactions with a final state consisting of a lepton, two ejected nucleons and an $(A-2)$ nucleus that is left with little or no excitation energy.
  The hammer events reported by the ArgoNeuT collaboration \cite{Acciarri:2014gev} were initially considered as detected events of that type. These events, however, have been shown to be related to pion production and reabsorption processes and not to exclusive $2N$ knockout \cite{Weinstein:2016inx}.
Other experiments using liquid argon detectors such as MicroBooNE \cite{microboone} and DUNE \cite{Adams:2013qkq} or scintillator trackers such as MINERvA \cite{minerva} and NOvA \cite{nova} are designed with the ability to observe $2N$ knockout events. 

  The exclusive $A(e,e' N_a N_b)$ cross section in the lab frame, can be written as a function of four response functions
  \begin{align}
    \dfrac{\mathrm{d}\sigma}{\mathrm{d}E_{e'} \mathrm{d}\Omega_{e'} \mathrm{d}T_a \mathrm{d}\Omega_a\mathrm{d}\Omega_b} = \sigma^\mathrm{Mott} g_{rec}^{-1} \nonumber \\
    \times \bigl[ v^e_{L} W_{CC} + v^e_T W_T + v^e_{TT} W_{TT} + v^e_{TL} W_{TC}  \bigr],\label{eq:electwo}
  \end{align}
  with recoil factor
  \begin{align}
    g_{rec} = \left| 1 + \frac{E_b}{E_{A-2}}\left(1 - \frac{ \vec{p}_b \cdot \left( \vec{q} - \vec{p}_a \right) }{p_b^2} \right) \right|.
  \end{align}
  Ten response functions contribute to $A(\nu_\mu,\mu^- N_a N_b)$ reactions
  \begin{align}
    \dfrac{\mathrm{d}\sigma}{\mathrm{d}E_\mu \mathrm{d}\Omega_\mu \mathrm{d}T_a \mathrm{d}\Omega_a\mathrm{d}\Omega_b} &= \sigma^{W} \zeta f_{rec}^{-1} \nonumber \\
    \times \bigl[ v_{CC} W_{CC} + v_{CL} W_{CL} &+ v_{LL} W_{LL} + v_T W_T \nonumber \\ 
      + v_{TT} W_{TT} + v_{TC} W_{TC} &+ v_{TL} W_{TL} \nonumber \\
    \mp (v_{T'} W_{T'} + v_{TC'} W_{TC'} &+ v_{TL'} W_{TL'}) \bigr].\label{eq:neuttwo}
  \end{align}
$T_a$ is the kinetic energy of particle $a$. The azimuthal information of the emitted nucleons is contained in $W_{TT}, W_{TC}, W_{TL}, W_{TC'}$ and $W_{TL'}$, while all the response functions depend on $\theta_a$ and $\theta_b$.

  In Fig.~\ref{fig:mecexcl} the results of an exclusive $^{12}$C$(\nu_\mu,\mu^- N_a N_b)$ calculation are shown for $2N$ knockout in the lepton-scattering plane. The top panel only includes the axial current of Eq.~(\ref{eq:axseaform3}). The panels 'sea' and 'pif' only use the vector seagull and pion-in-flight current respectively and the panel 'MECs' includes the coherent sum of vector and axial currents. The bottom panel shows the area where initial center-of-mass (c.o.m.)~momentum $\vec{P}_{12}$ of the pair,
\begin{align}
  \vec{P}_{12}= \vec{p}_a + \vec{p}_b - \vec{q},
  \label{eq:miss}
\end{align}
is smaller than 300 MeV/c.  

We observe that for the selected kinematic situation, the $2N$ knockout strength is dominated by the vector currents 'sea' and 'pif'. The strength of both currents is comparable in size. 
Further, the seagull and pion-in-flight currents interfere destructively, which can be inferred from the fourth panel. This destructive interference of the vector currents was also observed for exclusive \ce{^{16}O}$(\gamma,\p\n)$ interactions \cite{Ryckebusch:1993tf}. 
The $2N$ knockout strength is restricted to the part of phase space where the initial c.o.m.~momentum of the pair is kept low. To illustrate this, the region where $P_{12} < 300$ MeV/c is displayed in the bottom panel. 
The numerical results also show that the chosen kinematic situation favors back-to-back nucleon knockout in the lab frame, as studied e.g.~in \cite{Colle:2015ena}.
\subsection{Semi-exclusive $2N$ knockout}

\begin{figure}[ht!]
  \centering
  \includegraphics[height=0.8\textheight]{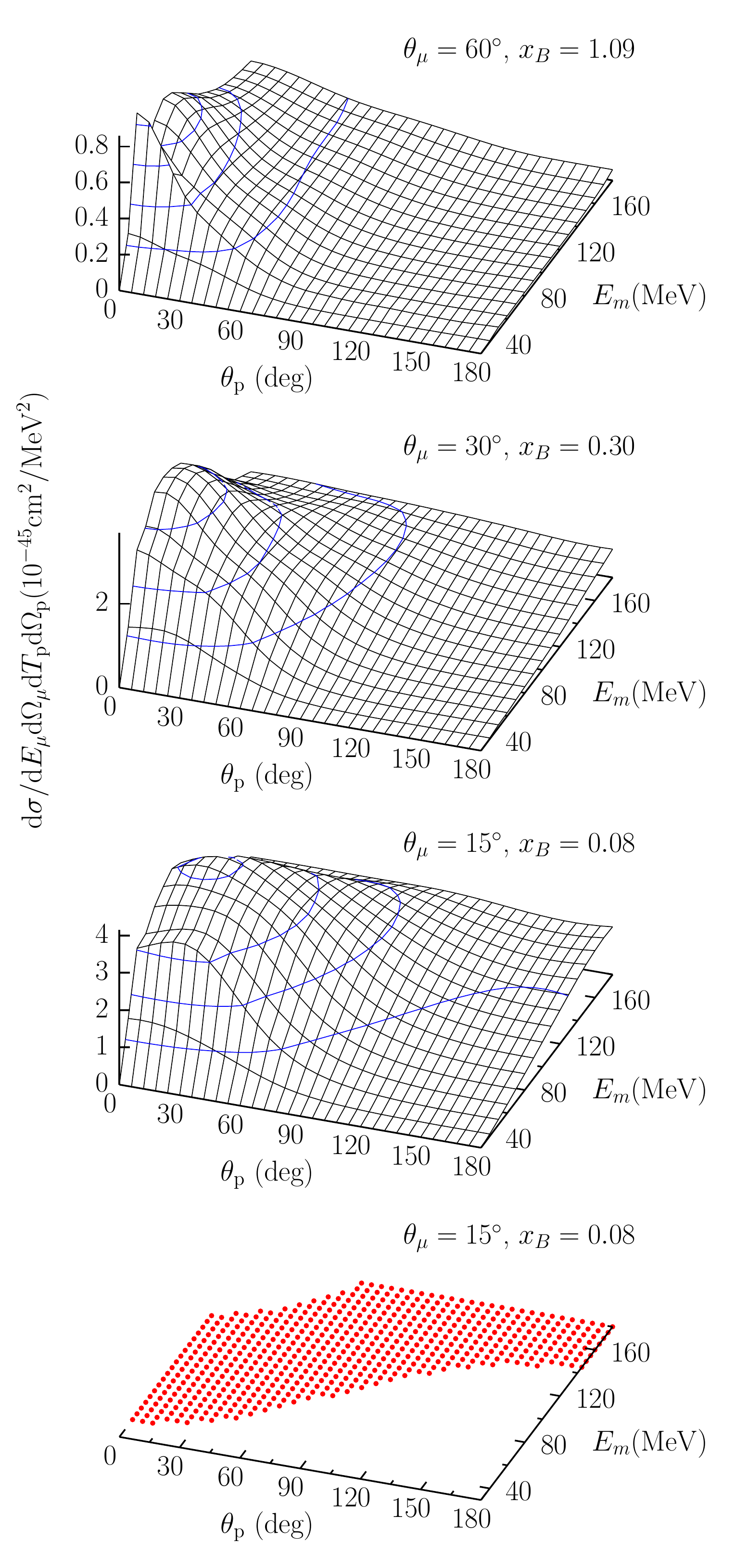}
  \caption{Semi-exclusive $^{12}$C$(\nu_\mu,\mu^-$p$)$ cross section for in-plane kinematics for $E_{\nu_\mu} = 750$ MeV, $E_\mu=550$ MeV and three muon scattering angles for. The bottom panel shows the ($\theta_\textnormal{p},E_m$) area with $P_{12} < 300$ MeV/c for $\theta_\mu=15^\circ$.}
  \label{fig:mecsemi}
\end{figure}

It is interesting to study the contribution of the exclusive $2N$ knockout $A(\nu_\mu,\mu^- N_a N_b)$ cross section to the inclusive $A(\nu_\mu,\mu^-)$ cross section, as there is very little data on exclusive cross sections. 

As an intermediate step, we compute the contribution of exclusive $2N$ knockout $A(l,l^\prime N_a N_b)$ strength to the $A(l,l^\prime N)$ cross section, where the residual nuclear system $(A-1)^*$ is excited above the $2N$ emission threshold. This is called the semi-exclusive cross section in this work. 

Exclusive $1N$ knockout cross sections detect the final state lepton and the emitted nucleon in coincidence. Processes with two emitted nucleons whereby one remains undetected also contribute to the signal.
This means that for neutrino experiments which have the ability to detect nucleons in the final state, but have a relatively high detection threshold, these semi-exclusive cross sections will be a very interesting tool.


The calculation of the semi-exclusive cross section involves an integration over the phase space of the undetected ejected nucleons. In the case where the detected particle is a proton, the total semi-exclusive cross section is a sum of the semi-exclusive $\p\p$ and $\p\n$ pair
knockout cross sections ($N_a =$ p, $N_b =$ p$^\prime$ or n),
\begin{align}
  &\dfrac{\mathrm{d}\sigma}{\mathrm{d}E_{l'} \mathrm{d}\Omega_{l'} \mathrm{d}T_\mathrm{p} \mathrm{d}\Omega_\textnormal{p}}(l,l' \mathrm{p}) \nonumber \\
  & \qquad = \int \mathrm{d}\Omega_\mathrm{p^\prime} \dfrac{\mathrm{d}\sigma}{\mathrm{d}E_{l'} \mathrm{d}\Omega_{l'} \mathrm{d}T_\mathrm{p} \mathrm{d}\Omega_\textnormal{p} \mathrm{d}\Omega_\mathrm{p^\prime}}(l,l' \mathrm{pp^\prime}) \nonumber \\ 
  & \qquad + \int \mathrm{d}\Omega_\textnormal{n} \dfrac{\mathrm{d}\sigma}{\mathrm{d}E_{l'} \mathrm{d}\Omega_{l'} \mathrm{d}T_\mathrm{p} \mathrm{d}\Omega_\textnormal{p} \mathrm{d}\Omega_\textnormal{n}}(l,l' \mathrm{pn})\label{eq:semi}.
\end{align}
We use the method outlined in \cite{Ryckebusch:1997gn} and exploit the fact that the exclusive $2N$ knockout strength resides in a well-defined part of phase space, see Fig.~\ref{fig:mecexcl}. In this limited part of the phase space, the momentum of the undetected particle $\vec{p}_b$ varies very little, which allows one to set $\vec{p}_b \approx \vec{p}_b^{\,ave}$. 
The average momentum ($\vec{p}_b^{\,ave}$) is determined by imposing quasi-deuteron kinematics ($P_{12} \approx 0$ in Eqn.~(\ref{eq:miss})
\begin{align}
  \vec{p}_b^{\,ave} = \vec{q} - \vec{p}_\mathrm{p}. 
\end{align}
With this average momentum, the integration over $\textnormal{d}\Omega_{\textnormal{p}^\prime}$ and $\textnormal{d}\Omega_{\textnormal{n}}$ in Eq.~(\ref{eq:semi}) can be performed analytically \cite{Ryckebusch:1997gn}

The results of a semi-exclusive \ce{^{12}C}$(\nu_\mu,\mu^- \p)$ calculation are displayed in Fig.~\ref{fig:mecsemi} for three different lepton scattering angles as a function of the outgoing angle of the detected proton $\theta_\p$ $(\phi_\p=0)$, and the missing energy $E_m = \omega - T_\p$.
The Bjorken variable $x_B=Q^2/2\omega m_N$ varies from 0.08 to 1.09 for the three presented kinematic situations.  
The semi-exclusive strength is largest for small $\theta_\mu$. Further, for large $\theta_\mu$, the strength is confined to small proton scattering angles, while relatively large strength at backward proton knockout is observed for small lepton scattering angles. 
This feature is related to the initial c.o.m.~momentum of the pair, the bottom panel shows the area where $P_{12} < 300$ MeV/c is accessible. This demonstrates that the semi-exclusive strength is dominated by pairs with small c.o.m.~momenta. 


 \subsection{Inclusive cross section results}\label{sec:incl}
 The $2N$ knockout contribution to the inclusive $A(l,l^\prime)$ cross section is calculated by integrating over the phase space $\textnormal{d}T_\textnormal{p} \textnormal{d}\Omega_\textnormal{p}$ in Eq.~(\ref{eq:semi})
 \begin{align}
   \frac{\mathrm{d}\sigma}{\mathrm{d}E_{l'}\mathrm{d}\Omega_{l'}} (l,l') = \int \mathrm{d}T_\textnormal{p} \mathrm{d}\Omega_\textnormal{p} \frac{\mathrm{d}\sigma}{\mathrm{d}E_{l'}\mathrm{d}\Omega_{l'} \mathrm{d}T_\textnormal{p}\mathrm{d}\Omega_\textnormal{p}} (l,l'\textnormal{p}).\label{eq:inclcross}
 \end{align}
 The angular integration can be done analytically, integration over the outgoing nucleon kinetic energy $T_\textnormal{p}$ is performed numerically.

 \begin{figure*}[ht!]
   \centering
   \includegraphics[width=0.95\textwidth,trim=0cm 0.6cm 0cm 0.3cm,clip]{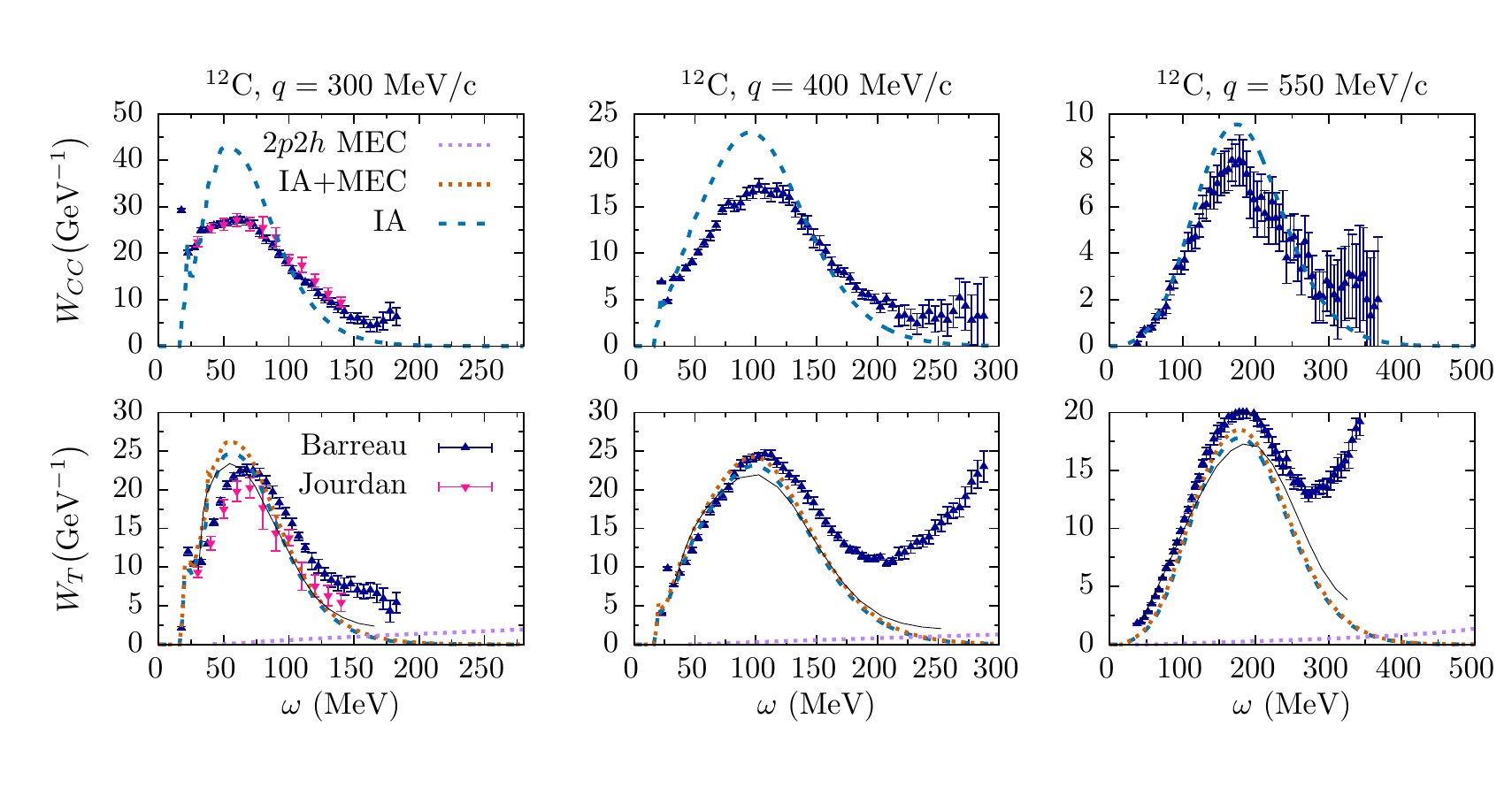}
   \caption{The 1p1h and 2p2h MEC response functions $W_{CC}$ and $W_T$ for $^{12}$C$(e,e^\prime)$ for three values of $q$. The dotted blue line is the contribution of the sea+pif currents. The solid black lines are the 1p1h+2p2h results from \cite{Amaro:1994fx}, the data is from Refs.~\cite{Barreau:1983ht,Jourdan:1996np}.}
   \label{fig:elecincl1p1h2p2hqmec}
 \end{figure*}

 \begin{figure*}[ht!]
   \centering
   \includegraphics[width=0.95\textwidth,trim=0cm 0.6cm 0cm 0.3cm,clip]{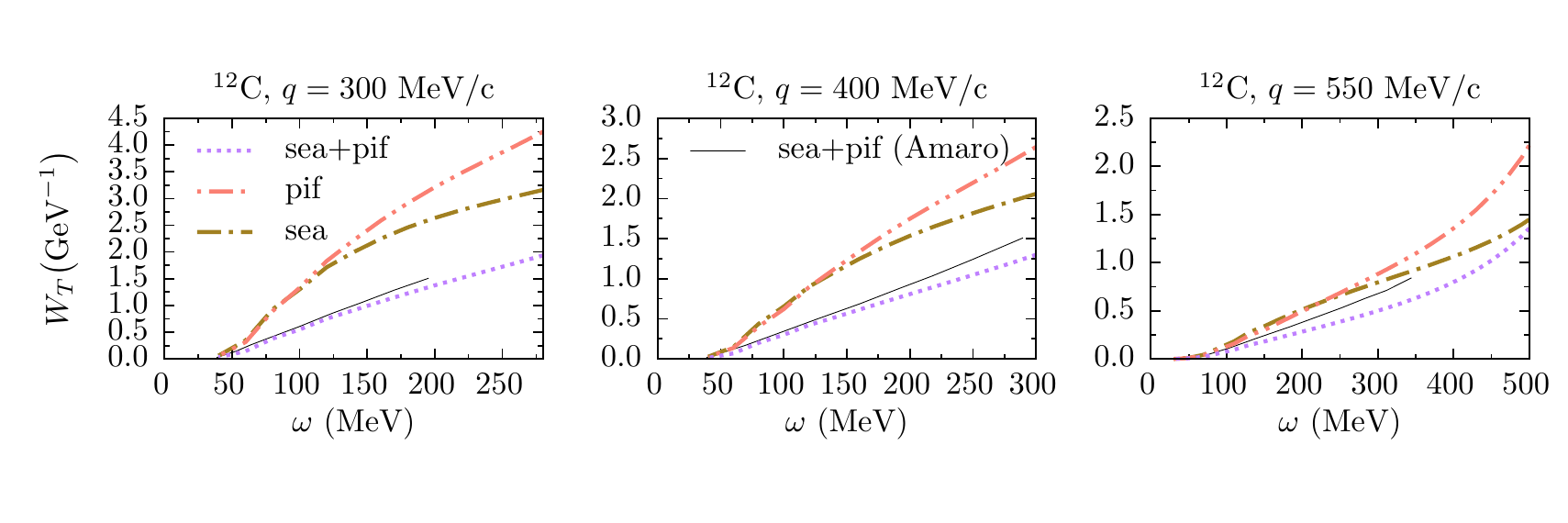}
   \caption{The 2p2h MEC response function $W_T$ for $^{12}$C$(e,e^\prime)$ for three values of $q$. The contributions of the seagull and pion-in-flight currents are shown separately. The solid black lines are the sea+pif results from \cite{Amaro:1994fx}.}
   \label{fig:elecincl2p2hqmec}
 \end{figure*}

\begin{figure}[h!]
  \centering
    \includegraphics[width=0.35\textwidth,trim=0cm 0.6cm 0cm 0.3cm,clip]{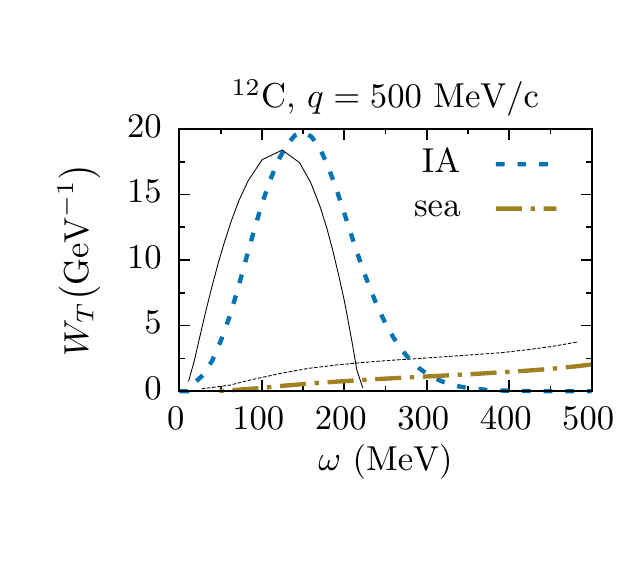}
    \caption{The transverse 1p1h and 2p2h responses for \ce{^{12}C}$(e,e^\prime)$ at $q=500$ MeV/c with only seagull currents. The solid (dashed) black lines are the 1p1h (2p2h) results from Ref.~\cite{Simo:2014aaa}.} 
  \label{fig:elec1p1h2p2hq500mec}
 \end{figure}

\begin{figure}[h!]
  \centering
    \includegraphics[width=0.35\textwidth,trim=0cm 0.6cm 0cm 0.3cm,clip]{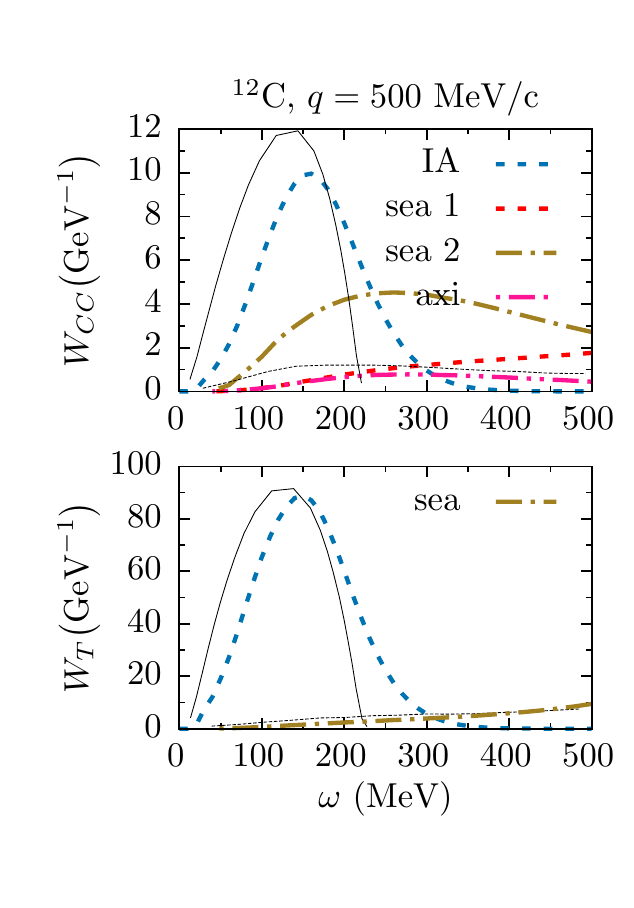}
    \caption{ The Coulomb and transverse 1p1h and 2p2h responses for \ce{^{12}C}$(\nu_\mu,\mu^-)$ at $q=500$ MeV/c with only seagull and axial currents. The solid (dashed) black lines are the 1p1h (2p2h) results from Ref.~\cite{Simo:2014aaa}.} 
  \label{fig:neut1p1h2p2hq500mec}
 \end{figure}

\begin{figure*}[h!]
  \centering
  \includegraphics[width=0.95\textwidth,trim=0cm 0.6cm 0cm 0.3cm,clip]{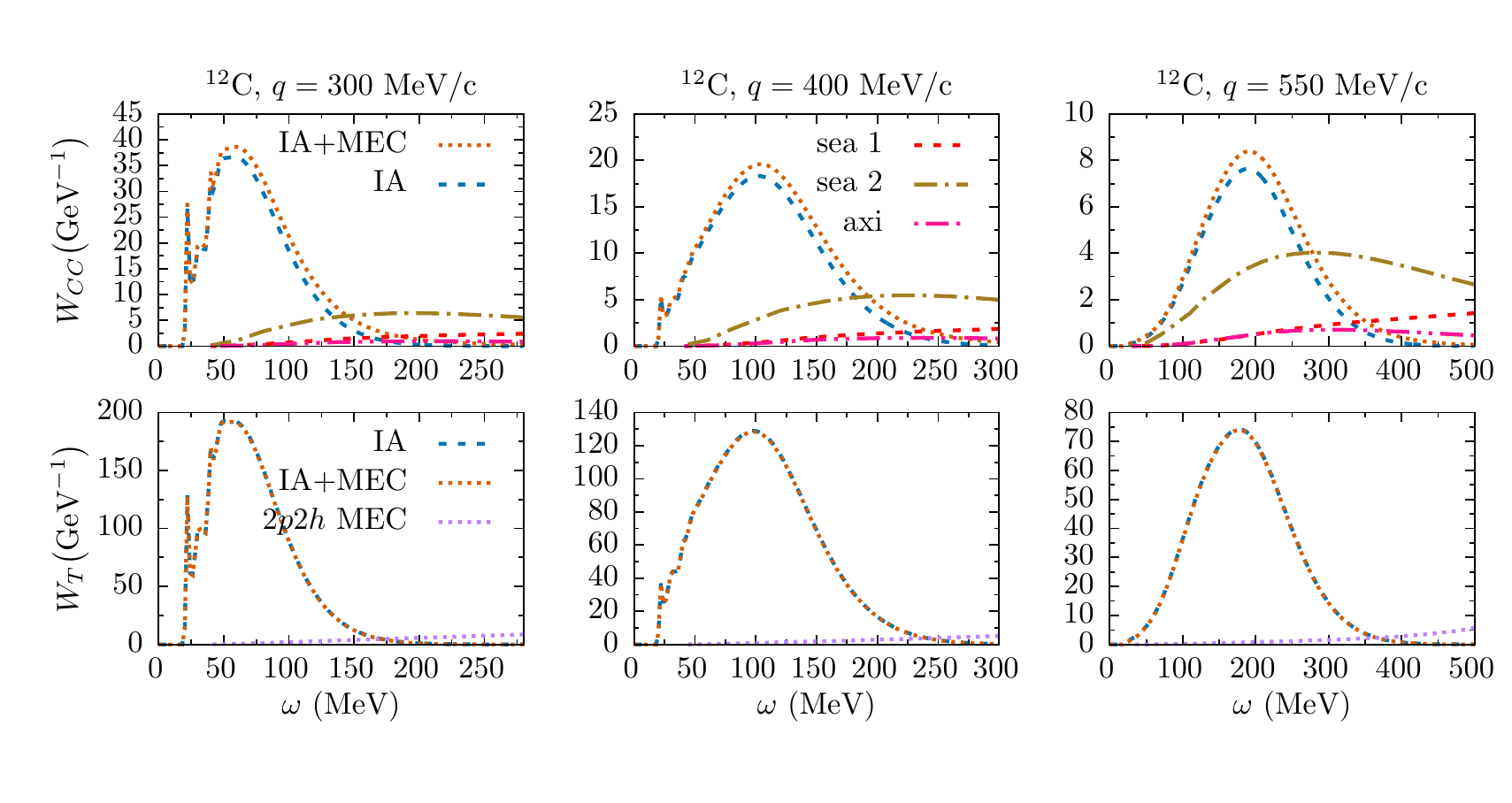}
  \caption{The 1p1h and 2p2h responses, same as Fig.~\ref{fig:elecincl1p1h2p2hqmec} but for $^{12}$C$(\nu_\mu,\mu^-)$.}
  \label{fig:neut1p1h2p2hqmec}
\end{figure*}

\begin{figure*}[h!]
  \centering
  \includegraphics[width=0.95\textwidth,trim=0cm 0.6cm 0cm 0.3cm,clip]{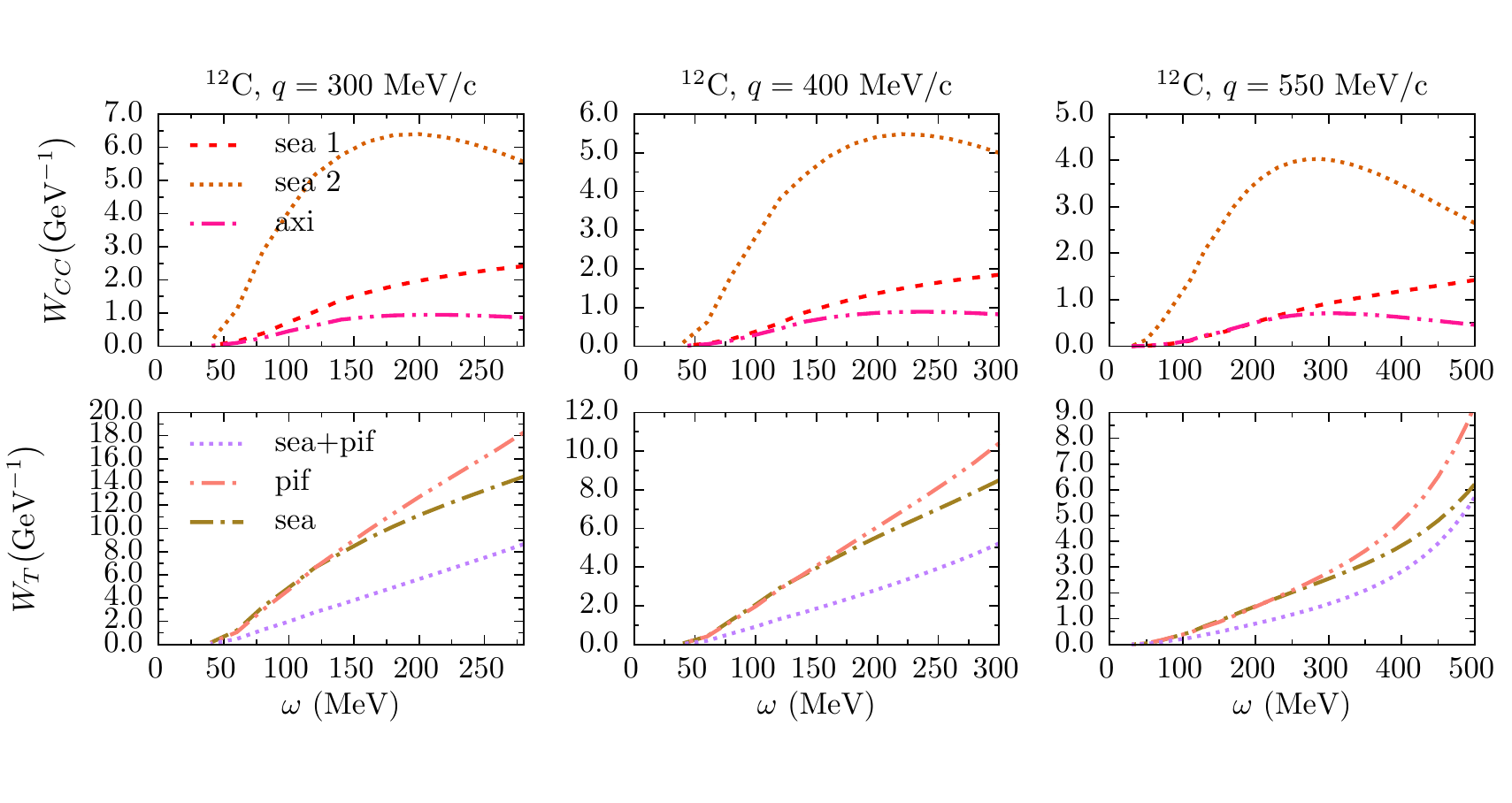}
  \caption{The 2p2h responses, same as Fig.~\ref{fig:elecincl2p2hqmec} but for $^{12}$C$(\nu_\mu,\mu^-)$ }
  \label{fig:neut2p2hqmec}
\end{figure*}

\begin{figure*}[h!]
  \centering
  \includegraphics[width=0.95\textwidth,trim=0cm 0.6cm 0cm 0.3cm,clip]{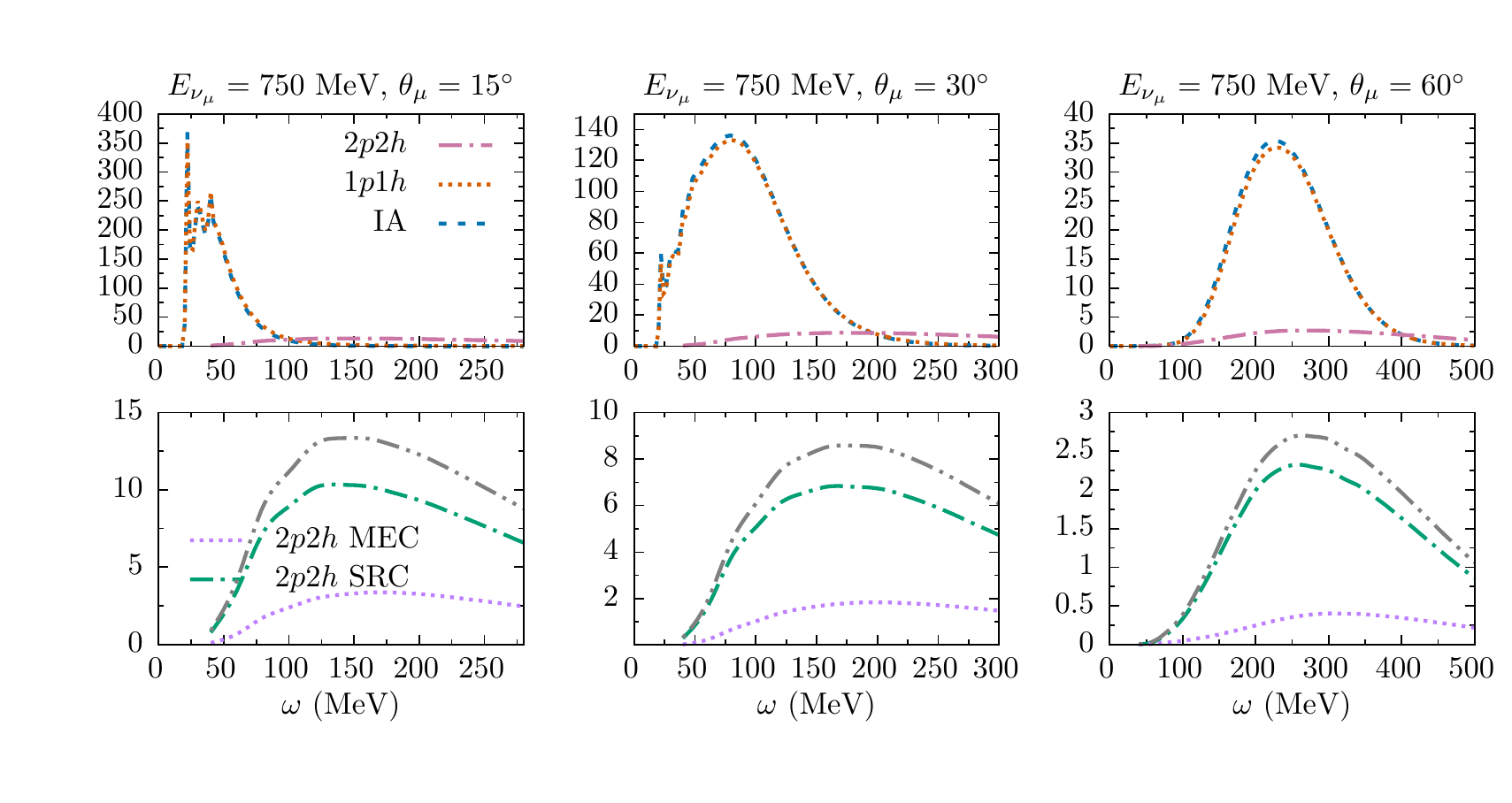}
  \caption{The $\omega$ dependence of the $^{12}$C$(\nu_\mu,\mu^-)$ cross section for $E_{\nu_\mu}=750$ MeV and three different values for the lepton scattering angle $\theta_\mu$. The top panels show the 1p1h and 2p2h cross sections. The bottom panels show the 2p2h part of the cross section, separating the contributions of SRCs and MECs.}
  \label{fig:neut1p1h2p2hmecsrc}
\end{figure*}

 Before we consider the inclusive $2N$ knockout responses for $\nu A$, we confront our results for electron scattering with data and other models \cite{Amaro:1994fx,Simo:2014aaa}.
 In Fig.~\ref{fig:elecincl1p1h2p2hqmec}, the 1p1h and 2p2h response functions $W_{CC}$ and $W_T$ are shown and compared with Rosenbluth separated cross-section data. The seagull and pion-in-flight currents have no effect on the Coulomb response, as the vector currents have no timelike component in the low-energy limit. In the 1p1h responses, the MECs result in a small increase of the responses. The 2p2h responses appear as a broad background to the 1p1h responses.
 Fig.~\ref{fig:elecincl2p2hqmec} shows the results of the inclusive transverse 2p2h responses, where the knockout of seagull and pion-in-flight pairs was studied separately. 
 The strength for both two-body currents is of similar size. The pion-in-flight current is slightly more important for large $\omega$ for the three-momentum transfers studied.
 More interesting is that the currents interfere destructively. In \cite{Amaro:2010iu} e.g, the same destructive interference was observed between the seagull and pion-in-flight currents in a relativistic Fermi gas model for $^{56}$Fe$(e,e^\prime)$.
 Further, our results are comparable to those of Amaro \textit{et al.}~\cite{Amaro:1994fx}, where a similar model was used. 

 In Figs.~\ref{fig:elec1p1h2p2hq500mec} and \ref{fig:neut1p1h2p2hq500mec}, inclusive $^{12}$C$(e,e^\prime)$ and $^{12}$C$(\nu_\mu,\mu^-)$ responses are studied at $q=500$ MeV/c. 
 Only the seagull current is accounted for in the $2N$ knockout calculations, to compare with the corresponding results of Ruiz Simo \textit{et al.}~\cite{Simo:2014aaa}, where a relativistic Fermi gas (RFG) was used. 
 The results of the $1N$ knockout calculations in the IA for both models are displayed as a reference.

 For electron scattering, the $2N$ knockout strength attributed to the seagull current is roughly a factor 2 smaller than in the RFG calculations of Ref.~\cite{Simo:2014aaa}. 
The $2N$ knockout contribution to the transverse response for $\nu A$ is very similar in both calculations. 
 In the Coulomb channel, the results for the three different axial currents are compared. The currents labeled 'sea,1' and 'axi' yield a strength that is comparable to each other and to the strength of the RFG calculations of Ref.~\cite{Simo:2014aaa}. For low $\omega$, the results of the currents 'sea,1' and 'axi' coincide. For increasing energy transfers, the former keeps increasing while the latter decreases for $\omega \gtrsim 250$ MeV. 
 The strength of the current 'sea,2', which was obtained after a nonrelativistic reduction of the axial seagull current in \cite{Simo:2014aaa}, is roughly five times larger than the other two prescriptions, and appears unrealistically large compared to the 1p1h strength.   

 The results for the responses $W_{CC}$ and $W_T$ for inclusive $^{12}$C$(\nu_\mu,\mu^-)$, including seagull and pion-in-flight currents in the 1p1h and 2p2h channels, are presented in Fig.~\ref{fig:neut1p1h2p2hqmec}. In the 1p1h channel, we only display the results using the 'axi' current. The results using the other two expressions can be inferred from Fig.~\ref{fig:neut1p1hqtwomec}. 
 In Fig.~\ref{fig:neut2p2hqmec}, the 2p2h responses are shown, showing the separate strengths of the seagull and pion-in-flight currents. 

 Comparing the 2p2h results in the transverse channel for electron and neutrino scattering, we observe that the contributions of the seagull and pion-in-flight currents have a similar $\omega$ dependence. The currents interfere destructively in both cases. The 2p2h responses for neutrino scattering are roughly a factor 4 larger than in electron interactions. The relative effect of the 2p2h responses in comparison with the 1p1h responses appears similar for electron and neutrino
 interactions. 
 The 2p2h Coulomb responses are smaller than the transverse responses, however their effect relative to the corresponding 1p1h response is larger. 

 In Fig.~\ref{fig:neut1p1h2p2hmecsrc}, the results of a $^{12}$C$(\nu_\mu,\mu^-)$ cross-section calculation are shown, for $E_{\nu_\mu} = 750$ MeV and three muon scattering angles. In the calculations the SRCs are accounted for, as outlined in \cite{VanCuyck:2016fab}, next to the MECs, including interference. 
 The effect of the MECs on the 1p1h channel is negligible, as can be inferred from Fig.~\ref{fig:neut1p1h2p2hqmec}. The double differential cross sections are dominated by the transverse channel and the effect of the MECs on the transverse responses is very small. The decrease of the 1p1h channel due to the presence of two-body currents is mainly caused by SRCs, as shown in \cite{VanCuyck:2016fab}.

 The contribution of the MECs in the 2p2h channel yields a smaller contribution to the inclusive cross section than that provided by the SRCs. It is roughly a factor 3 smaller for $\theta_\mu = 15^\circ$ to a factor 5 for $\theta_\mu = 60^\circ$. 
 The results suggest that the total 2p2h strength equals the sum of the SRC and MEC contributions, however a small destructive interference is present between both types of two-body currents.
 The combined effect of both types of two-body currents yields strength that appears as a broad background to the QE peak, ranging from the $2N$ knockout threshold into the dip region. 
 In the dip region, experimental data is underpredicted by calculations in the IA, and the $2N$ knockout contribution provided by SRC and MEC pairs only accounts for a small fraction the missing strength in this region.

 \section{Flux-folded double differential cross sections}\label{sec:flu}
 

 In Ref.~\cite{Pandey:2016jju}, the impact of long-range correlations on $\nu A$ cross sections was studied in a CRPA approach. Flux-folded double differential cross sections off $^{12}$C were presented and compared with MiniBooNE and T2K data. The CRPA model underpredicted the data because of the absence of processes beyond pure QE scattering. 
 These calculations are extended with $2N$ knockout of MEC and SRC pairs. 

\begin{figure*}[ht!]
  \centering
    \includegraphics[width=0.80\textwidth,trim=0cm 0cm 0cm 0.0cm,clip]{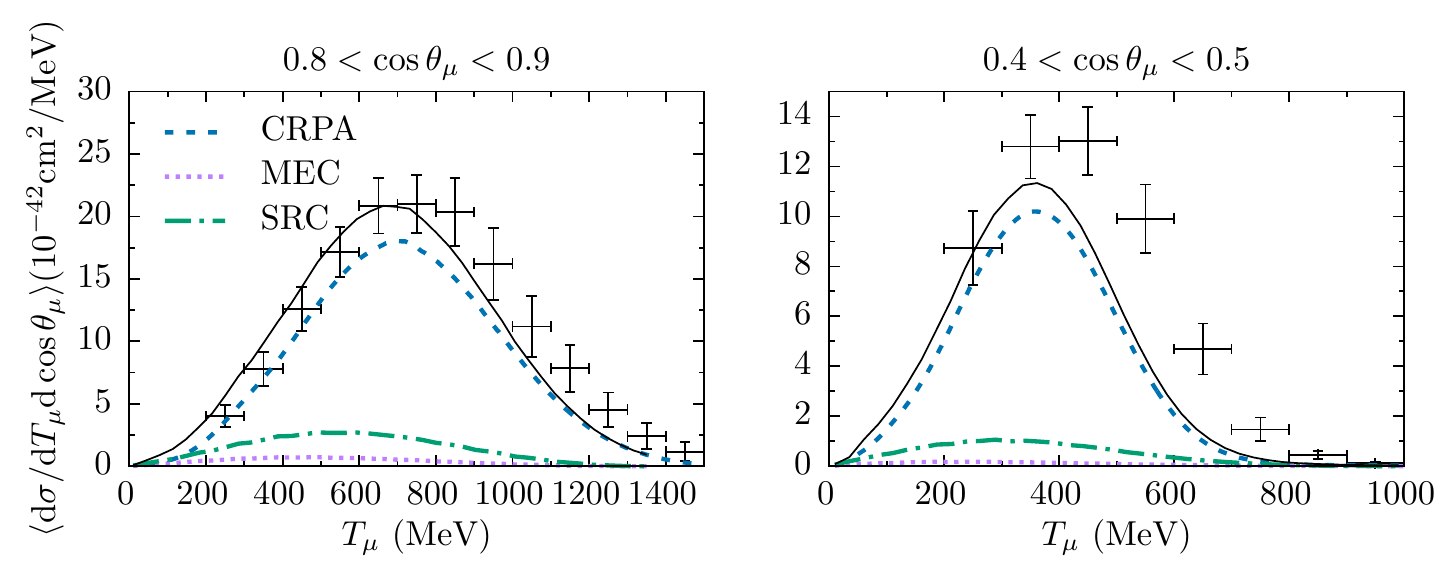}
    \caption{MiniBooNE CCQE-like flux-folded double differential cross sections per target neutron for \ce{^{12}C}$(\nu_\mu,\mu^-)$. The solid line is the sum of the three contributions. Data is from Ref.~\cite{AguilarArevalo:2010zc}, the experimental error bars represent the shape uncertainties, flux uncertainties are not included.} 
  \label{fig:neutminiboone}
\end{figure*}

 In Fig.~\ref{fig:neutminiboone}, a prediction for the strength of the MiniBooNE flux-folded differential CCQE-like cross section is shown as a function of the muon kinetic energy $T_\mu$, and
compared with data. 
The solid black line is the sum of the different contributions. 
Due to the heavy computational cost, the flux-folding was done in steps of 100 MeV, while the integration in $\cos \theta_\mu$ was done in three steps. 


The MiniBooNE CCQE-like data set is defined as the processes where one muon and no pions are observed in the final state. Yet, in the analysis, the CCQE-like data has partly been corrected for $\Delta$-currents by subtracting pion-less $\Delta$ decays from the data \cite{teppei}.  

The CRPA approach combined with $2N$ knockout of MEC and SRC pairs reproduces the strength and shape of the forward bin in Fig.~\ref{fig:neutminiboone} very well, however, the predictions and data appear to be shifted over some 50 MeV. 
The agreement in the bin with more backward lepton scattering is less satisfactory, as a large fraction of the measured strength is not accounted for by the calculations.

\begin{figure*}[ht!]
  \centering
  \subfloat[y][Inclusive T2K data from Ref.~\cite{Abe:2013jth}.]{\includegraphics[width=0.80\textwidth,trim=0cm 0cm 0cm 0.0cm,clip]{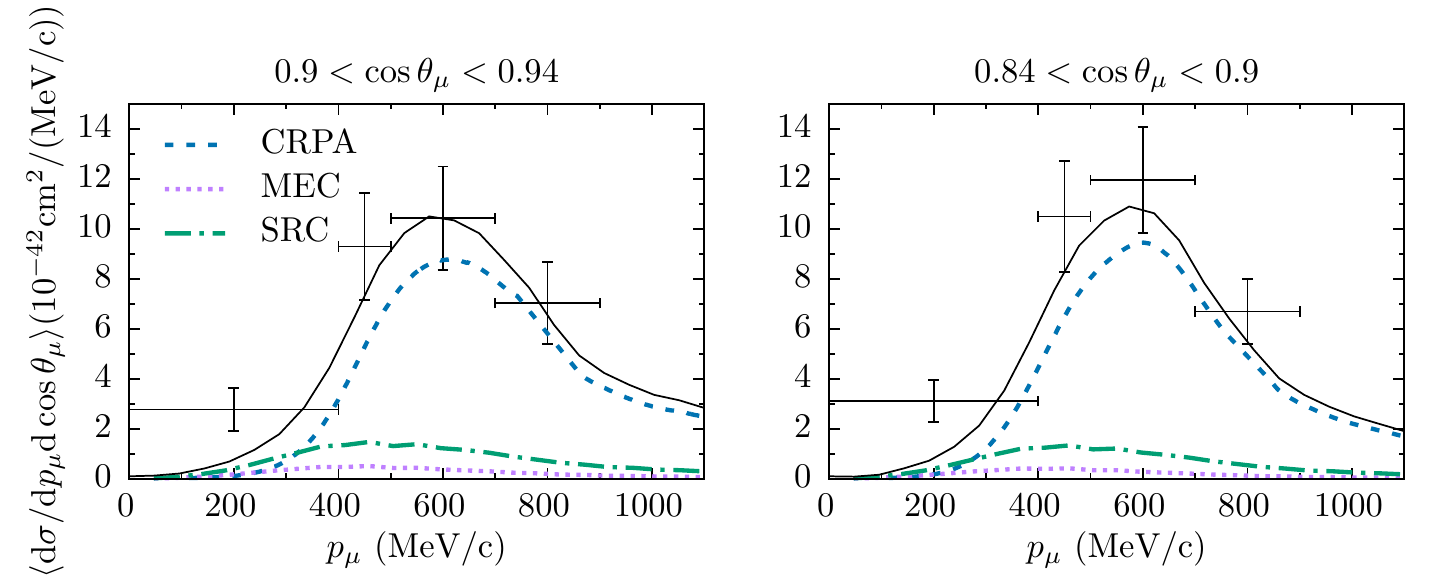}}
    \label{fig:neutt2kincl}
    \subfloat[z][CC0$\pi$ T2K data from Ref.~\cite{Abe:2016tmq}.]{\includegraphics[width=0.80\textwidth,trim=0cm 0cm 0cm 0.0cm,clip]{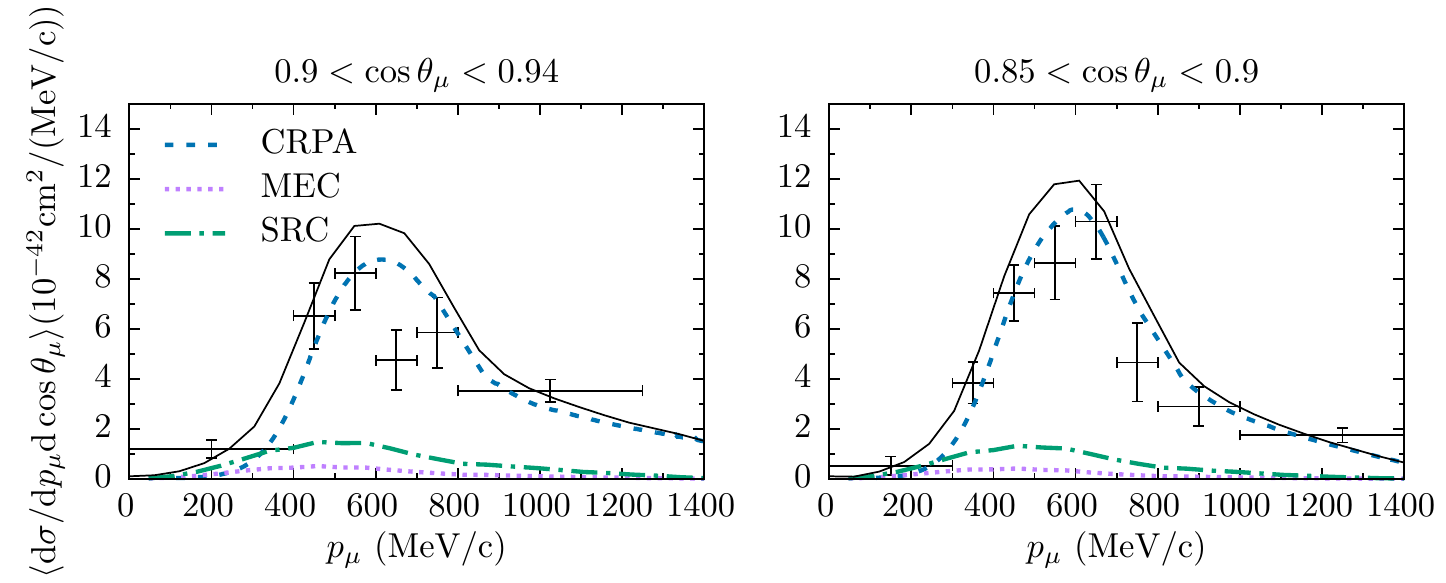}}
  \label{fig:neutt2kccqe}
    \caption{T2K flux-folded double differential cross sections per target nucleon for \ce{^{12}C}$(\nu_\mu,\mu^-)$. The solid line is the sum of the three contributions. The experimental error bars represent the shape uncertainties, flux uncertainties are not included.} 
  \label{fig:neutt2k}
 \end{figure*}

In Fig.~\ref{fig:neutt2k}, the corresponding double differential results for two T2K angular bins are shown as a function of the muon momentum $p_\mu$. Two bins were used for the averaging over $\cos \theta_\mu$. 
In the top panels, the results are compared with the inclusive data, i.e.~processes with pions in the final state are included.
In the bottom panels, the results are compared with T2K CC0$\pi$ data, defined as the processes where no pions are observed in the final state. This data was not corrected for the $\Delta$-current contribution, and they should be included in the $2N$ knockout channel for a complete description of the data. 

The theoretical predictions reproduce the inclusive data rather well, while extra strength from $\Delta$-currents and pion production can be included without overestimating the data. 
The prediction of the CC0$\pi$ data appears to be on the high side, as the $\Delta$-current contribution has not been included. Yet, a satisfactory description of the data is not ruled out as interference effects between SRCs, MECs and $\Delta$-currents should be included, and the flux normalization error of the data should be accounted for. 

The results of the flux-folded double differential cross sections are in line with the unfolded cross sections displayed in Fig.~\ref{fig:neut1p1h2p2hmecsrc}. The strength of the SRCs is a factor 3 to 5 larger than that of the MECs.

 \section{Summary}\label{sec:sum}

In this work we studied the effect of seagull and pion-in-flight currents on $\nu A$ cross sections. The research presented here is part of a larger project studying multinucleon effects on $\nu A$ interactions. The presented formalism provides a framework for the treatment of MECs and SRCs in the calculation of exclusive, semi-exclusive and inclusive $1N$ and $2N$ knockout cross sections.

The standard expressions for the vector seagull and pion-in-flight currents were used. 
For the axial current, three different prescriptions are used. The first version is constructed by multiplying the axial seagull current by $G_A(Q^2)$. The second expression followed after a nonrelativistic reduction of the axial seagull current in \cite{Simo:2016ikv}. The third expression was derived in \cite{Towner:1992np}, where a two-nucleon version of the PCAC hypothesis was used to constrain the current. 

The second expression of the axial seagull current yields an unrealistically large $2N$ knockout contribution to the inclusive double differential cross section. This unrealistic behavior might be related to the pion-pole current. This current was not taken into account in this work, but in \cite{Simo:2016ikv} it was included to fulfill the PCAC relation.
For small $\omega$, the first expression of the axial seagull current and the current labeled 'axi' result in a similar increase of the responses in the $1N$ knockout channel and give rise to comparable $2N$ knockout strength. For larger $\omega$, the current labeled 'sea,1' consistently yields more strength. 
The axial current 'axi' fits most naturally in the model presented in this work, since it fulfills the two-body version of the PCAC hypothesis. 
We will be guided by the conclusions drawn from this research for the inclusion of axial $\Delta$-currents in future work, which are generally assumed to provide the largest strength. 

The inclusion of MECs in double differential cross section calculations of electron and neutrino scattering interactions yields relatively small effects, as the various terms
tend to cancel each other. 
The inclusion of seagull and pion-in-flight currents in electron scattering interactions results in a small increase of the 1p1h channel and a broad background 2p2h strength. 
For neutrino scattering, the combined effect of the MECs on the 1p1h channel is very small. 
The $2N$ knockout strength appears as a background to the 1p1h channel, extending into the dip region where the data is severely underpredicted by the IA. The MECs account for only a small fraction of the missing strength.

Accounting for long-range correlations in the $1N$ knockout channel in a CRPA approach and MECs and SRCs in the $2N$ knockout channel, a fair agreement with the MiniBooNE CCQE-like data is reached in the bin $0.8 < \cos \theta_\mu < 0.9$. For $0.4 < \cos \theta_\mu < 0.5$, some strength is missing. 
A fair agreement with the T2K data is reached. Taking interference effects and the additional flux normalization uncertainty into account, there is room for the extra strength from $\Delta$-currents and pion production. 

The results presented here used \ce{^{12}C} as a target nucleus, but the model is general and can be used for all target nuclei with a $0^+$ ground state such as \ce{^{16}O} and \ce{^{40}Ar}. 


 \appendix*

 \section{Matrix elements}\label{app:mat}
 The standard expressions for the multipole operators and the nuclear currents are used in this work, see e.g. Refs.~\cite{walecka2004theoretical,VanCuyck:2016fab}.
 The 2p2h matrix elements for the vector part of the seagull and pion-in-flight currents are given in \cite{Ryckebusch:1993tf}. The matrix elements for the three axial currents of Eqs.~(\ref{eq:axseaform1}-\ref{eq:axseaform3}) are given by
 
 \begin{widetext}

  \begin{align}
    \langle ab;J_1 & \parallel \widehat{M}_J^{\rm{Coul}} \left[ \widehat{\rho}_A^{\,[1],\rm{sea,1}}(1,2) \right] \parallel cd;J_2 \rangle = \frac{1}{g_A} \left( \frac{f_{\pi NN}}{m_\pi} \right)^2 \frac{1}{\sqrt{4\pi}}  \frac{2}{\pi} G_A(Q^2) \sum_l \sum_{J_3} \sum_{\eta=\pm 1} \widehat{J} \widehat{J}_1 \widehat{J}_2 \widehat{J}_3 (-1)^{J_3+l} \nonumber \\
    & \times  \bra{ab} \vec{I}_V \ket{cd} \sqrt{l+\delta_{\eta,+1}}\threej{J,l,J_3,0,0,0} 
    \int \d p \frac{p^3}{p^2+m_\pi^2} \Gamma_\pi^2(p^2)
    \int \d r_1 \int \d r_2 \nonumber \\
    \times \Vast( & \langle j_a \parallel j_J(q r_1) j_l(p r_1) Y_{J_3}(\Omega_1) \parallel j_c \rangle_{r_1} \langle j_b \parallel j_{l+\eta}(p r_2) \left[ Y_{l+\eta}(\Omega_2) \otimes \vec{\sigma}_2 \right]_l \parallel j_d \rangle_{r_2} \ninej{j_a,j_b,J_1,j_c,j_d,J_2,J_3,l,J} \nonumber \\
    - (-1)^{l+J_3+J} & \langle j_a \parallel  j_{l+\eta}(p r_1) \left[ Y_{l+\eta}(\Omega_1) \otimes \vec{\sigma}_1 \right]_l \parallel j_c \rangle_{r_1} \langle j_b \parallel j_{J}(q r_2) j_l(p r_2) Y_{J_3}(\Omega_2) \parallel j_d \rangle_{r_2} \ninej{j_a,j_b,J_1,j_c,j_d,J_2,l,J_3,J} \Vast),
    \label{eq:seaax1}
  \end{align}

  \begin{align}
    \langle ab;J_1 & \parallel \widehat{M}_J^{\rm{Coul}} \left[ \widehat{\rho}_A^{\,[2],\rm{sea,2}}(1,2) \right] \parallel cd;J_2 \rangle = \frac{1}{g_A} \left( \frac{f_{\pi NN}}{m_\pi} \right)^2 \frac{1}{\sqrt{4\pi}}  \left( \frac{2}{\pi} \right)^2  \sum_{l l^\prime} \sum_{\eta=\pm 1} \widehat{J} \widehat{J}_1 \widehat{J}_2 \widehat{l} (-1)^{l-l^\prime} \nonumber \\
    & \times  \bra{ab} \vec{I}_V \ket{cd} \sqrt{l^\prime+\delta_{\eta,+1}}\threej{l,l^\prime,J,0,0,0} \nonumber \\
    &\times \int \d p_1 \, p_1^2 F_\pi(p_1^2) 
    \int \d p_2 \frac{p_2^3}{p_2^2+m_\pi^2} \Gamma_\pi^2(p_2^2)
    \int \d r_1 \int \d r_2 \int \d r r^2 j_l(p_1 r) j_{l^\prime}(p_2 r) j_J(qr) \nonumber \\
    \times \Vast( & \langle j_a \parallel j_l(p_1 r_1) Y_{l}(\Omega_1) \parallel j_c \rangle_{r_1} \langle j_b \parallel j_{l^\prime+\eta}(p_2 r_2) \left[ Y_{l^\prime+\eta}(\Omega_2) \otimes \vec{\sigma}_2 \right]_{l^\prime} \parallel j_d \rangle_{r_2} \ninej{j_a,j_b,J_1,j_c,j_d,J_2,l,l^\prime,J} \nonumber \\
    - (-1)^{l+l^\prime+J} & \langle j_a \parallel  j_{l^\prime+\eta}(p_2 r_1) \left[ Y_{l^\prime+\eta}(\Omega_1) \otimes \vec{\sigma}_1 \right]_{l^\prime} \parallel j_c \rangle_{r_1} \langle j_b \parallel  j_l(p_1 r_2) Y_{l}(\Omega_2) \parallel j_d \rangle_{r_2} \ninej{j_a,j_b,J_1,j_c,j_d,J_2,l^\prime,l,J} \Vast).
    \label{seaax2}
  \end{align}
 \end{widetext}
  In the matrix elements, we used the shorthand notation $a \equiv (n_a,l_a,1/2,j_a)$. The radial transition densities $\langle a || \widehat{\mathcal{O}}_J || b \rangle_r$ are defined such that they are related to the full matrix elements as $\langle a || \widehat{\mathcal{O}}_J || b \rangle \equiv \int \textnormal{d} r \langle a || \widehat{\mathcal{O}}_J || b \rangle_r$.
 The matrix element for the axial current 'axi' is obtained by removing the $G_A(Q^2)$ in Eq.~(\ref{eq:seaax1}) and introducing the $p$-dependent form factor $F_\pi(p^2)$. 

 \begin{acknowledgments}
   This work was supported by the Interuniversity Attraction Poles Programme P7/12 initiated by the Belgian Science Policy Office and the Research Foundation Flanders (FWO-Flanders). The authors thank T. Katori for clarification of the MiniBooNE data. The computational resources (Stevin Supercomputer Infrastructure) and services used in this work were provided by Ghent University, the Hercules Foundation and the Flemish Government. 
 \end{acknowledgments}

 \bibliography{biblio}

 \end{document}